\def\puncspace{\ifmmode\,\else{\ifcat.\C{\if.\C\else\if,\C\else\if?\C\else%
\if:\C\else\if;\C\else\if-\C\else\if)\C\else\if/\C\else\if]\C\else\if'\C%
\else\space\fi\fi\fi\fi\fi\fi\fi\fi\fi\fi}%
\else\if\empty\C\else\if\space\C\else\space\fi\fi\fi}\fi}
\def\SP{\let\\=\empty\futurelet\C\puncspace}

\def\h1{$h^{-1}$\SP}

\def\ie{{\it i.e.\/}\rm\ }
\def\eg{{\it e.g.\/}\rm\ }
\def\lsim{~\rlap{$<$}{\lower 1.0ex\hbox{$\sim$}}}
\def\gsim{~\rlap{$>$}{\lower 1.0ex\hbox{$\sim$}}}
\def\void#1{{}}

\def\xi2{$\chi^2$}

\documentclass{aa}
\usepackage{graphicx,psfig,amsmath}
\usepackage{rotating}

\begin{document}

%

   \author{ E. Hatziminaoglou  \inst{1} 
 \and M.A.T. Groenewegen \inst{1}
 \and L. da Costa        \inst{1}
 \and S. Arnouts         \inst{1}
 \and C. Benoist         \inst{2}
 \and R. Madejsky        \inst{1,3}
 \and R. P. Mignani      \inst{1}
 \and L. F. Olsen	 \inst{4}
 \and C. Rit\'e          \inst{1,5}
 \and M. Schirmer        \inst{6}
 \and R. Slijkhuis       \inst{1}
 \and B. Vandame         \inst{1} 
          }

\institute{
     European Southern Observatory, Karl-Schwarzschild-Str. 2, 
     D-85748 Garching b. M\"unchen, Germany
\and Observatoire de la C\^ote d'Azur, BP 229, 06304 Nice, cedex 4, France
\and Universidade Estadual de Feira de Santana, Campus
     Universit\'ario, Feira de Santana, BA, Brazil
\and Astronomical Observatory, Juliane Maries Vej 30, DK-2100 Copenhagen, Denmark
\and Observat\'orio Nacional, Rua Gen. Jos\'e Cristino 77, 
     Rio de Janerio, R.J., Brazil 
\and Max-Planck Institut f\"ur Astrophysik, Karl-Schwarzschild-Str. 1,
     D-85748 Garching b. M\"unchen, Germany
}

\titlerunning{Exploring the point-source catalogues of CDF-S}
\authorrunning{Hatziminaoglou et al.}

\title{ESO Imaging Survey}{\subtitle{Exploring the optical/infrared
imaging data of CDF-S: Point Sources} 

\abstract{This paper describes the methodology currently being implemented in
the EIS pipeline for analysing optical/infrared multi-colour data. The
aim is to identify different classes of objects as well as possible
undesirable features associated with the construction of colour
catalogues.  The classification method used is based on the
$\chi^2$-fitting of template spectra to the observed SEDs, as measured
through broad-band filters. Its main advantage is the simultaneous use
of all colours, properly weighted by the photometric errors. In
addition, it provides basic information on the properties of the
classified objects (\eg redshift, effective temperature). These
characteristics make the
\xi2-technique ideal for handling large multi-band
datasets. The results are compared to the more traditional
colour-colour selection and, whenever possible, to model
predictions. In order to identify objects with odd colours, either
associated with rare populations or to possible problems in the
catalogue, outliers are searched for in the multi-dimensional colour
space using a nearest-neighbour criterion. Outliers with large
\xi2-values are individually inspected to further investigate 
their nature. The tools developed are used for a preliminary analysis
of the multi-colour point source catalogue constructed from the
optical/infrared imaging data obtained for the Chandra Deep Field
South (CDF-S). These data are publicly available, representing the
first installment of the ongoing EIS Deep Public Survey.
}

   \date{Received 14 August 2001, accepted}
   \offprints{Evanthia Hatziminaoglou,
   \email{ehatzimi@eso.org}}

   \maketitle


\section{Introduction}
\label{intro}

The ESO Imaging Survey (EIS) is an ongoing project to carry out public
imaging surveys in support of VLT.  Its primary goal is to provide
multi-wavelength data sets from which samples comprising different
types of extragalactic and galactic objects can be extracted for
follow-up spectroscopic observations. So far the surveys conducted
have used different instrument/telescope setups to carry out
moderately deep observations of large areas, deep optical/infrared
observations of high-galactic latitude fields, contiguous areas of the
SMC and LMC and selected stellar fields including open clusters, and
globular clusters (for more details see da Costa, 2001).  Altogether
over 50~square~degrees of the southern sky have already been surveyed,
albeit using different filter combinations and reaching different
magnitudes (see the EIS web page at
``http://www.hq.eso.org/science/eis/'').

The ultimate success of these surveys will, to a large extent, depend
on the ability of reliably identifying different classes of objects
and extracting well-defined samples for spectroscopic follow-up
observations. While colour selection is nothing new and several
methods have been devised and applied in the past, the demands of
modern, wide-area surveys involving large numbers of objects and
passbands is relatively new and must be properly addressed.
Therefore, to fully achieve the scientific goals of EIS a detailed
understanding of the distribution of objects in colour space is
required. This is not only necessary for the selection of
spectroscopic targets but also as a verification of the colour
catalogues being routinely produced by the survey pipeline.

An ideal way of tackling this problem is to combine intrinsic (\eg
spectral properties) and statistical information (\ie spatial
distribution, luminosity and mass functions, evolution) regarding
different classes of known objects. The nature of objects, as
characterised by its spectral properties, can be assessed by comparing
the measurements obtained from the multi-colour photometry with those
estimated using template spectra describing different types of
objects. To take into account the statistical properties of a given
population requires detailed simulations of the stellar population of
our Galaxy as well as the extragalactic populations. These simulations
must also satisfy observational constraints, such as sky position,
completeness, photometric errors, and morphological classification. In
principle, combining these two independent methods should lead to a
further improvement of the classification of objects extracted from
colour catalogues. As a first step towards this goal, this paper
discusses the classification of the objects based exclusively on their
spectral properties as derived from multi-colour observations. As a
practical illustration, this analysis is applied to the multi-colour
point source catalogues extracted from the recently completed
optical/infrared data of the Chandra Deep Field South (CDF-S) by the
EIS Deep Public Survey (DPS; Vandame et al., 2001; Arnouts et al.,
2001a).

The $UBVRI$ optical data covers an area of 0.25~square~degrees. These
data are complemented by $JK_s$ near-infrared observations over
0.1~square~degrees located at the centre of the area covered by the
optical data. While the angular coverage is relatively small, this is
the first complete data set of this survey which at the end will cover a
total area of 3~square~degrees corresponding to 12 times the data
presented here.  Therefore the results presented in this paper provide
a first assessment of the likely outcome of this survey once
completed. Using the CDF-S data as a benchmark is particularly
interesting considering the large number of imaging and spectroscopic
observations planned for this region, in addition to the already
publicly available deep X-ray observations of Chandra (Rosati et al.,
2001). These observations will provide an unprecedented
multi-wavelength data set that should certainly help refine the
classification algorithms being developed.

In Section~\ref{data} the data as well as the method employed in the
construction of the point source catalogue are briefly
reviewed. Section~\ref{methods} presents the methods used to classify
objects based on their colour properties and to search for objects
located in poorly populated regions of the colour space. These methods
are applied to the CDF-S optical and optical/infrared data and the
results are discussed and compared to other methods in
Section~\ref{Results}. In this section tables listing different types
of objects are also presented. In Section~\ref{evaluation} an
assessment of the results is carried out by visually inspecting image
cutouts and examining the photometric measurements of individual
objects to evaluate the performance of catalogue production and target
selection procedures. General conclusions and a discussion of the main
results is presented in Section~\ref{discussion}. Finally, in
Section~\ref{summary} a brief summary is presented.

\section{Colour Catalogues}
\label{data}

The first step for a target selection from multi-colour data is the
construction of multi-colour catalogues.  Such catalogues can
be produced either by association of objects listed in the single
passband catalogues or by building a $\chi^2$ image (Szalay et al., 1999)
combining the
available images in the different passbands. The reference $\chi^2$
image is used to detect objects, while their photometric measurements
are carried out in the individual single passband images. For the
work presented here the first technique has been applied. The $\chi^2$
image will be used in future work where a detailed discussion of the
pros and cons of the two methods will be presented.

The input to the catalogue association are the single passband
catalogues (Vandame et al., 2001; Arnouts et al., 2001a) cut at a
$S/N=2$, slightly lower than the catalogues publicly released which
were cut at a $S/N$=3.  Furthermore, in the present work only objects
inside a trimmed region and with appropriate SExtractor flags are
considered. Finally, the $U$-band catalogue used here was extracted
from an image produced stacking all the available $U/38$ and $U350/60$
images, in order to reach a fainter limiting magnitude. A more
detailed discussion of this catalogue and of its photometric
calibration will be present in Arnouts et al. (2001b).

After the association, only objects that are in the area common to all
catalogues and outside all the masks placed around saturated objects
and bright stars, are kept. This ensures that all objects could have
been detected in all passbands. An object is included in the final
colour catalogue if it is detected with a $S/N\ge$3 in at least one
passband. This was done as a compromise between completeness, to avoid
pruning possible interesting objects before looking at the data, and
the number of spurious detections.  If an object is not detected in a
particular passband, its magnitude is set to the corresponding
3$\sigma$ limit. Note that the 3$\sigma$ limiting magnitudes
(throughout this work all magnitudes are in the Vega system) of the
colour catalogues are: $U$=25.7, $B$=27.4, $V$=26.0, $R$=25.9,
$I$=24.7, $J$=23.3, and $K_s$=21.3. 

In the single passband catalogue, point and extended sources are
separated using the SExtractor CLASS\_STAR flag, up to the
morphological classification limit. The point/extended source
classification in a colour catalogue is not trivial and the following
scheme is adopted. As the point/extended source classification works
best for bright objects, the passband utilised for the classification
is the one where the object is brightest with respect to the
classification limit in that filter (if any). If the object is
classified as a point source in this filter, it is considered a point
source in the colour catalogue. This procedure is valid as long as the
seeing on the images obtained in different passbands is comparable,
which is the case here.

The final point-source catalogues considered comprise 1539 and 623
objects in five and seven passbands, respectively. Based on empirical
number counts (Wolf et al., 2001a) the completeness of the point-source
catalogue is estimated to be $\ge 90$\% at $R$=24.5. The present
analysis only considers sub-samples consisting of objects detected in
at least three bands and those consisting of objects detected
only in the red-most passband available over the two regions covered
in five and seven filters.

\section{Analysis}
\label{methods}

\subsection{Object Classification}

Over the years several techniques have been employed to classify
objects based on their colour properties, the most common being the
selection of objects within regions defined in two-dimensional
projections of colour space. These regions are defined based on colour
tracks computed from template spectra.  While this is a reasonable
approach for data sets encompassing a small number of passbands, it is
cumbersome to handle photometric errors and as the number of bands
increases the problem rapidly becomes untrackable.  Furthermore, use
of a colour-colour diagram inevitably leads to severe contamination,
requires the use of several projections in order to properly
constraint a particular population and is not easily implemented for
automatic classification for large datasets.

A more suitable approach to handle colour data from large imaging
surveys is the $\chi^2$-technique (\eg Wolf et al. 2001b; 2001c)
originally developed for galaxy
photometric redshift estimation (\eg Bolzonella et al., 2000 and
references therein) and quasar search (Hatziminaoglou et al., 2000,
hereafter HMP00; Richards et al., 2001 and references therein). 
This method is an optimal way of simultaneously
handling samples consisting of more than three passbands, reducing the
dimensionality of the problem. It consists of
fitting the observed spectral energy distribution (SED) of each object
to a series of template spectra of different classes of objects,
minimising the $\chi^2$ given by

\begin{equation}
\chi^2= \sum_{i=1}^n \frac{(F_{obs}^i-F_{mod}^i(z))^2} {\sigma_i^2}
\end{equation}
\noindent
$F_{obs}^i$ and $F_{mod}^i$ are the observed and the template fluxes
in the $i$ band, respectively, and $\sigma_i$ is the estimated error
of the observed flux in this band. The error model adopted includes
the error in the magnitudes as estimated by SExtractor added in
quadrature to the zero-point error, determined in the photometric
calibration of each band. Since SExtractor tends to underestimate the
errors for faint sources, their amplitude was corrected to match that
estimated from comparisons with external data (\eg Arnouts et al.,
2001a). The above equation is normalised to the passband with the
smallest photometric error, although other ways are possible. This
subtracts one degree of freedom from the \xi2-analysis.

The SED of each object is compared to all available templates
separately, and the object is assigned a given type depending on the
value of the \xi2. Classifications are considered robust (rank 1) if
the \xi2 is within 95\% confidence level; good (rank 2) if in the
interval 95-99\% and poor (rank 3) if outside the 99\% confidence
level. Poorly classified objects can indicate an incomplete spectral
library, errors in the (theoretical) spectra, inadequacies in the
error model adopted and/or undesirable features in the colour
catalogue. The reliability of the results, of course, critically
depend on the completeness and the quality of the available spectral
library. The classification can always be refined by continuously
adding measured or improved theoretical spectra. It is important to
point out that even if the number of filters is limited to three this
approach, in principle equivalent to the traditional colour-colour
selection, provides a more convenient way of handling the errors and
uses additional constraints based on flux upper-limits.

The spectral library currently in use consists of series of: model
quasar, white dwarf and brown dwarf spectra; three empirical cool
white dwarf spectra; a set of stellar templates; and a set of model
galaxy spectra.  The template quasar spectra are assumed to have three
different power-law continua (spectral indexes of 0, 0.5 and 1 in the
optical).  The emission lines (Ly$_\alpha$, Ly$_\beta$, CIII, CIV,
MgII, SiIV, H$_\alpha$, H$_\beta$, and H$_\gamma$) are assumed to have
Gaussian profiles and typical relative intensities and equivalent
widths (\eg Peterson, 1997).  The Ly$_\alpha$ forest has been modeled
according to Madau (1995). Template spectra were created for redshifts
from $z=0$ to $z=6$, in steps of d$z=0.1$.  A set of theoretical white
dwarf spectra (provided by D. K\"oster; for the description of the
models see Finley et al., 1997; Homeier et al., 1998), with effective
temperatures and effective gravities ranging from 6000K to 100000K and
$\log g = 7 - 9$, respectively, has been incorporated.  Furthermore,
three empirical spectra of very cool white dwarves ($T_{\rm eff} <
4000$K) have been included, two covering the optical wavelength range
(Ibata et al., 2000), and one covering both the optical and
near-infrared wavelength range (Oppenheimer et al., 2001).  Also
included is a series of theoretical low mass stars and brown dwarf
spectra (Chabrier et al., 2000) covering temperatures from 500K to
2800K corresponding roughly to masses in the range 0.03 to 0.1
M$_{\odot}$, for an effective gravity of $\log g=4.5$. The effective
temperature range corresponds approximately to spectral types later
than M6V (Kirkpatrick et al., 1991). We should point out that the
transition between main sequence M-dwarves and low mass stars is
arbitrary, since there is an overlap of templates of objects with
temperatures around 2800K. Stellar spectra are taken from the stellar
library of Pickles (1998), which contains 131 spectra for a broad
range of stellar types (main sequence, giant and sub-giant
stars). Finally, the set of galaxy spectra used is the Coleman, Wu \&
Weedman (1980) model, with no intrinsic evolution. This last set of
templates, however, and since this paper only deals with point
sources, is only used in cases of poor classifications, as it will be
explained later on.  The conversion from spectra to magnitudes are
carried out using the same response functions as the observations
(Vandame et al., 2001; Arnouts et al., 2001a).

It is worth pointing out that a classification scheme based on the
\xi2-technique may lead to degeneracies, which are type-dependent (\eg
quasars - galaxies - stars). Multiple minima may occur in the
parameter space (\eg redshift - type) and such degeneracies
can only be solved by including additional information in the classification
procedure. This will be addressed in a forthcoming paper (Hatziminaoglou
et al., 2002).  In the present work only objects with real detections
in at least three passbands are treated. One could also use upper limits
as an {\it a posteriori}, in order to impose additional constraints and 
exclude highly improbable solutions. An alternative way is to construct
realistic mock catalogues and to combine the colour-space
distribution of the different populations with the individual properties 
(\ie SED) of the objects (\eg Hatziminaoglou et al., 2002).

\subsection{Searching for Outliers}
\label{searchout}

The technique presented above assumes that all objects belong to
classes for which the spectral properties are known.  In order not to
overlook the presence of unknown populations and to identify possible
problems in the process of building the colour sample, it is of great
interest to be able to pick up colour-space ``outliers'', even though
these may simply belong to rare populations, with known spectral
properties.

For the systematic identification of outliers the nearest neighbour
criterion is adopted using two dissimilarity measures (\eg Warren et
al., 1991). The first measure is the Euclidean distance in colour space
given by

\begin{equation}
d_{ii',S1}=\sum_{j=1}^N (c_{ij}-c_{i'j})^2,
\end{equation}
\noindent
The second is this distance weighted by the photometric errors:
 
\begin{equation}
d_{ii',S2}=\sum_{j=1}^N (c_{ij}-c_{i'j})^2/(\sigma^2_{c_{ij}}
+\sigma^2_{c_{i'j}})
\end{equation}
where $N$ is the number of possible colours derived from $n$ filters
($N=\frac{n!}{2!(n-2)!}$), $c_{ij}$ the colour $j$ of object i and
$\sigma_{c_{ij}}$ the error in the colour. A ``nearest-neighbour''
criterion is applied next and for each object the $m-th$ smallest
value of the respective $d_{ii',S1}$ and $d_{ii',S2}$ arrays is taken
as the degree of isolation of the object from its neighbours (\eg the
stellar locus).  In this paper only the cases $m=2$, to search for
individual isolated objects, and $m=3$, as an example to isolate
different types of objects, are considered. Depending on the
characteristics of the survey it might be desirable to use higher
values of $m$ to identify possible grouping of rare objects, as
originally discussed by Warren et al. (1991). Following these authors,
a selection criterion is applied depending on the position of the
objects in the d$_{S_1}$ versus d$_{S_2}$ plane, which isolates
objects with peculiar colour properties from the rest of the
sample. Alternative cluster analysis algorithms are also being
examined and will be presented elsewhere.

\section{Results}
\label{Results}

\subsection{Quasar Candidates}
\label{quasar}

From the original sample of 1539 objects, morphologically classified
as point sources (see Section~\ref{data}) within the
0.25~square~degrees covered in five optical passbands, 1494, detected
in at least three filters, were considered.  In total there are 204
objects classified as quasars in the magnitude range $19\lsim
B\lsim25$. This number is in excellent agreement with that predicted
by quasar models (202$^{+131}_{-78}$,
\eg Hatziminaoglou et al., 2002) at the limiting magnitude of
$B$=25.0, which roughly corresponds to the completeness limit of the
colour catalogue considered. It is also in good agreement with
estimates based on observed number counts (\eg Hartwick \& Schade
1990; Glazebrook et al., 1995).

\begin{figure*}[ht]
\centerline{
\psfig{figure=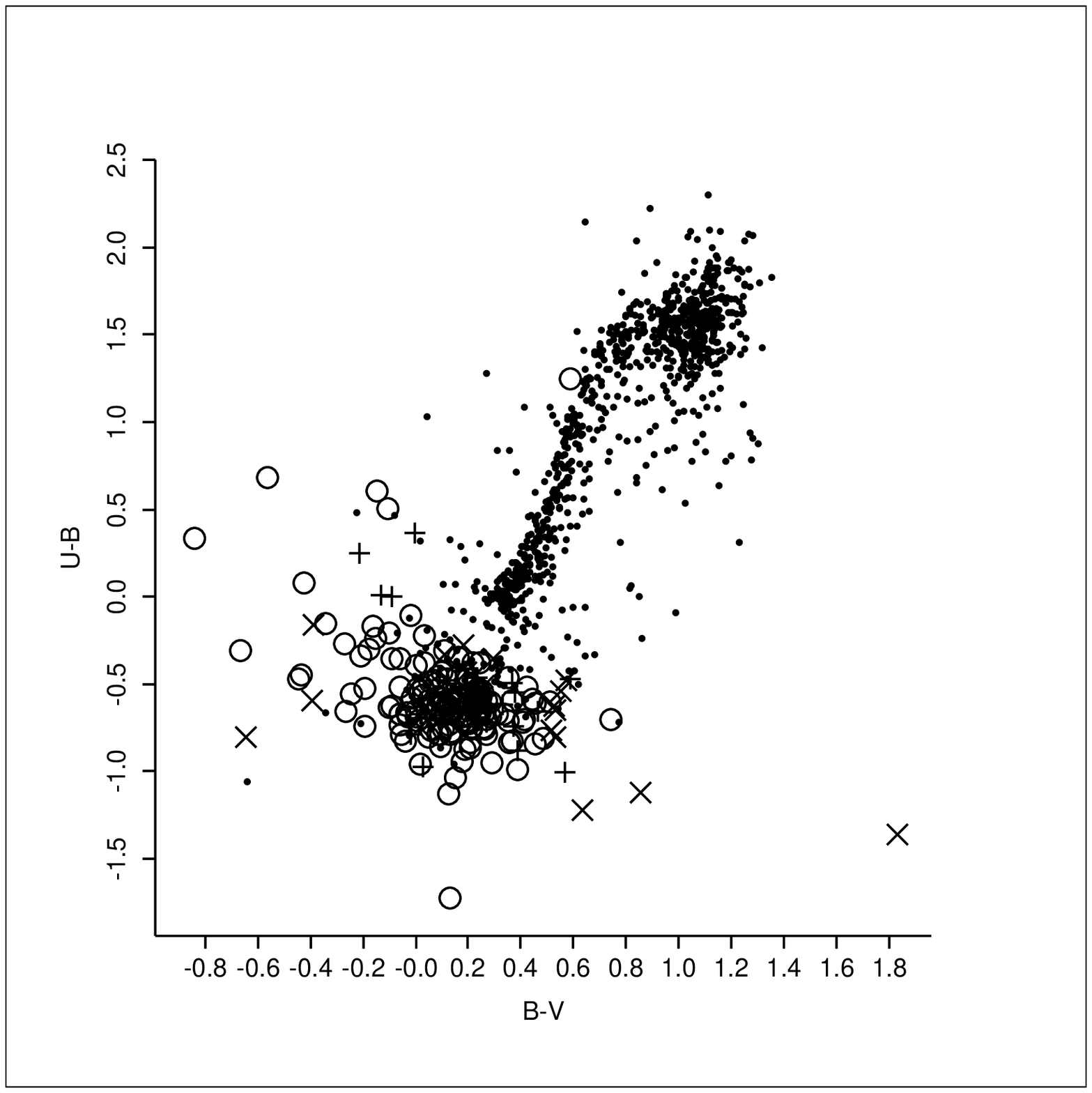,height=8cm,width=8cm}
\psfig{figure=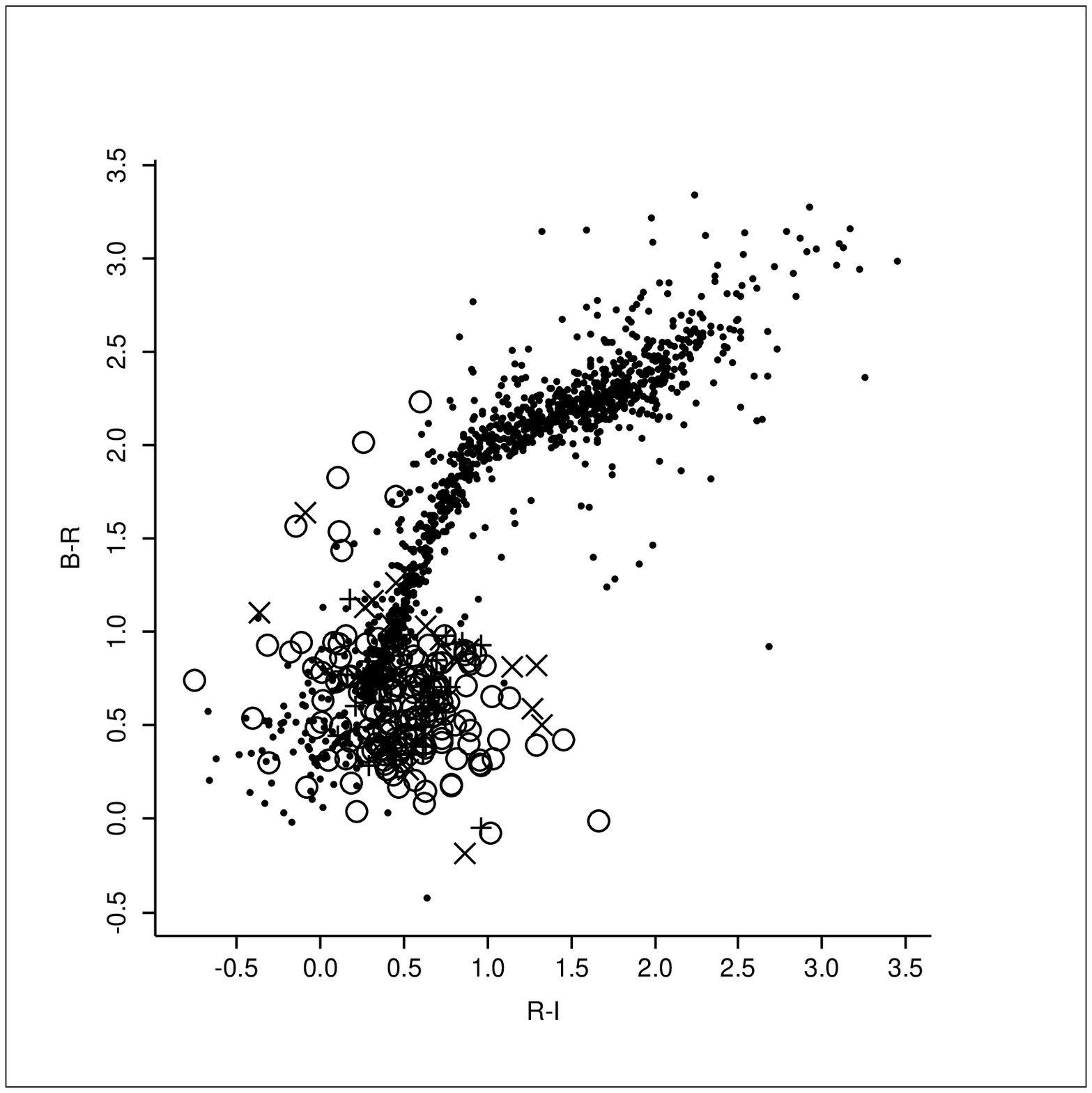,height=8cm,width=8cm}}
\caption{$(U-B)/(B-V)$ (panel (a)) and $(B-R)/(R-I)$ (panel (b))
two-colour diagrams showing the quasar candidates selected on the
basis of the \xi2-technique. Open circles, plus signs and crosses
denote candidates classified at 95\%, 95-99\% and outside 99\%
confidence level, respectively.}
\label{fig:UBBVqso}
\end{figure*}

To help evaluate the results of the \xi2-method it is important to
compare them with those obtained from a simple colour-colour
selection.  This is illustrated in Figure~\ref{fig:UBBVqso} which
shows the two colour-colour diagrams normally used to identify low to
intermediate ($(U-B)/(B-V)$) and high-redshift quasars
($(B-R)/(R-I)$), respectively. In the plot the identified quasars are
displayed with different symbols to indicate their associated
confidence level as follows: robust (open circles); good (plus signs);
and poor (crosses). The same notation applies to all colour-colour
plots presented hereafter, unless noted otherwise.

From the careful inspection of these colour-colour diagrams several
points can be made regarding the performance of the
\xi2-analysis. Note, for instance, that nearly all robust candidates are
located in the region predicted by the models for low- to
intermediate-redshift ($z\lsim2.2$) quasars,
\ie $-0.4\lsim(B-V)\lsim0.7$ and $-1.2\lsim(U-B)\lsim-0.3$, with
a few cases extending into the region where higher redshift quasars
are expected to lie.  One important short-coming of the standard
colour selection is its inability to discriminate between objects with
similar colours, leading inevitably to contamination problems. In the
particular case of $z\lsim2.2$ quasars, the main contaminants are
white dwarves and early spectral type main-sequence stars. The results
of \xi2-analysis suggest that this technique is capable of
distinguishing among these different classes, as can been seen from
quasar candidates lying very close to and sometimes overlapping
objects of other types. Poor candidates are, in general, located at
the outskirts of the region delineated by the robust candidates,
except for a few cases where the two populations overlap. Similar
conclusions can be drawn from the $(B-R)/(R-I)$ diagram shown in panel
(b). In particular, given the $B$ filter used, high-redshift quasars
($z\gsim3.0$) lie very close to the main sequence, making a colour
selection problematic and susceptible to contamination by
main-sequence stars. 
It is worth mentioning the object with
$B-V~\sim1.8$ (see panel (a)) located in a region unlikely to be
occupied by quasars. However, on other colour diagrams this object
lies close to quasar tracks. A more detailed discussion about its
nature will be presented in Section~\ref{evaluation}.

In order to evaluate the results obtained applying the
\xi2-method, they can be compared with those based on the more
traditional UVX and BRX selection. Adopting criteria similar to those
used by Hall et al. (1996) one finds 298 UVX and 49 BRX independent
quasar candidates. Out of these, 166 are in common with those
classified using the \xi2-method. Considering the
\xi2-classification as reference one estimates a contamination 
of $\sim40\%$ in a colour-colour quasar selection, demonstrating the
potentiality of the technique for minimising the number of
contaminants.  Moreover, since it uses the colour information in a
combined way, it should also lead to a higher completeness than those
based on distances from the stellar locus (\eg Gaidos, Magnier \&
Schechter 1993; Newberg \& Yanny 1997).  For example, colour based
selections will tend to miss quasars with redshifts in the range
$2.5\lsim z\lsim3.5$. This is a serious drawback because currently it
is believed that the space density of optically-selected quasars
starts decreasing in this redshift range. The possible benefits of the 
\xi2-classification over a simple colour selection remains to be 
evaluated, when spectroscopic data become available.

\begin{figure}
\centerline{
\psfig{figure=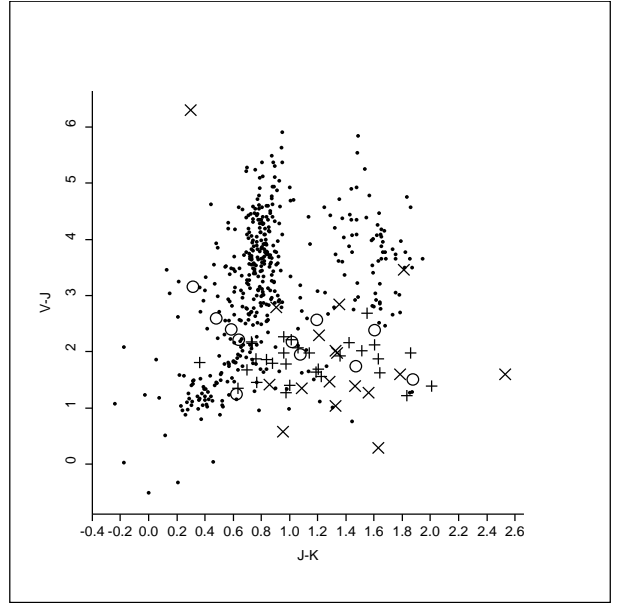,height=8cm,width=8cm}}
\caption{Optical-infrared  colour-colour diagram used in the selection
of KX candidates, showing the quasar candidates selected on the basis
of the $\chi^2$-technique. The symbols correspond to those defined in
Figure~\ref{fig:UBBVqso}.}
\label{fig:kxqso}
\end{figure}

The CDF-S field has also been observed in the near-infrared $JK_s$
passbands over an area of $\sim~0.1$~square~degrees.  For this seven
passband sub-sample, the number of morphologically classified point
sources is 623, out of which 605 are detected in at least three
filters. In total there are 92 objects classified as quasar
candidates, out of which 62 are in common with those found using only
the optical data. The infrared information yields 30 new
candidates. Five cases originally classified as quasars are now
assigned different classifications: three as white dwarves, one
as an G8I and one as a K3I stars.

For the case of optical/infrared all \xi2-selected quasar candidates
are shown on the $(V-J)/(J-K_s)$ diagram, introduced by Croom et
al. (2001), presented in Figure~\ref{fig:kxqso}.  This diagram is
suitable for identifying quasar candidates, due to the following
reasons. First, it could partially solve the degeneracy between low to
intermediate quasars and white dwarves, occuring when the UVX
criterion is applied. Quasars should exhibit a K-excess due to the
broad bump created by dust, present in their spectra at the
wavelengths around $\sim 1\mu$m (in the rest frame). White dwarves do
not have such near-infrared spectral features and should be separated
from quasars when the infrared information is added. Second, as
pointed out by Croom et al. (2001), KX-selection could identify
reddened quasars possibly missed by the UVX selection. Choosing
similar regions of colour space as these authors, one finds 65 quasar
candidates, 52 of which in common with those UVX selected. Of the
KX-selected candidates 34 belong to the quasar candidate list defined
using the \xi2-technique.  From Figure~\ref{fig:kxqso} one sees that
in order to ensure completeness of \xi2-selected robust quasar
candidates, one is forced to consider all objects with $V-J<3.0$,
which in turn leads to a contamination comparable ($\sim$ 50\%) to the
one introduced by the UVX selection. Another point worth mentioning is
the fact that in Figure~\ref{fig:kxqso} one finds red objects not
predicted by models of the spectral properties of point-sources.  As
discussed below, this is probably due to the contamination of the
sample by unresolved galaxies (see Section~\ref{evaluation}).  This
population is seen in all optical/infrared colour-diagrams presented
below.

Another important feature of the the \xi2-method is that it can be
used not only to classify the objects but also, in the case of
extragalactic candidates, to estimate their redshifts. The photometric
redshift distribution estimated for the quasar candidate sample
selected from the five optical passbands by the \xi2-technique is
shown in Figure~\ref{fig:zphot}, where the three rankings are plotted
separately. The distribution covers a broad range of estimated
redshifts extending all the way to $z\sim4.5$. The present sample
includes 16 candidates with estimated redshifts $z\gsim3.5$, among
which 10 are robust classifications.  Due to the degeneracy in the
assignment of quasar redshifts based solely on broad-band optical
filters (HMP00) objects with $z\lsim2$ have a considerable dispersion
in their estimated photometric redshifts. This accounts for the excess
seen at $z\lsim0.5$ and the dearth at $1\lsim z\lsim2$.  At $z\sim1$
there is also the increased probability of misclassifying Compact
Emission Line Galaxies (CELGs) as quasar candidates, as shown by HMP00
using the DMS spectroscopic sample. The dearth of objects with
redshifts in the interval $2.5\lsim z\lsim3.5$ is due to the colours
of AGN at these redshifts, which are much like the colours of the main
sequence stars, and can be very easily mis-classified as such.  Note
that in the present sample there are seven $U$-dropouts and a $B$-dropout
robust candidates with photometric redshifts $z\gsim~3.5$.

\begin{figure}
\centerline{
\psfig{figure=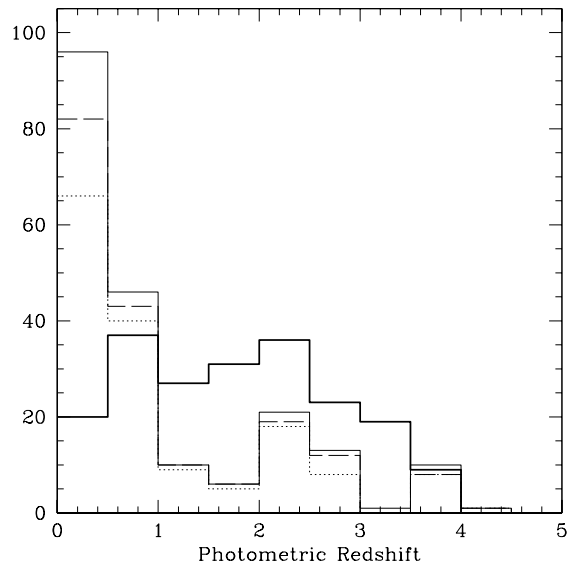,width=8cm,height=8cm}}
\caption{Photometric redshift distribution of quasar candidates
selected by the minimum $\chi^2$-method for the $UBVRI$ data set. The
solid histogram denotes the distribution for candidates selected when no 
constraint is applied on the values of the \xi2. The
dotted histogram corresponds to objects selected at 99\%; and the
dashed histogram correspond to those selected at 95\%, as described in
the text. Finally, the thick solid histogram is that predicted according the
model adopted for the evolution of the quasar LF.}
\label{fig:zphot}
\end{figure}

The redshift distribution of the 92 quasar candidates selected from
their optical/infrared photometry is presented in
Figure~\ref{fig:zphot7}. A comparison between Figures \ref{fig:zphot7} and
\ref{fig:zphot} shows that by including the infrared data one can
significantly improve the redshift estimates. As can be seen, the
excess of low-redshift quasars is considerably smaller and the
distribution resembles more closely the model prediction in the
redshift interval $2 \lsim z \lsim 3$. For the objects in common,
Figure~\ref{fig:z7vsz5} shows the comparison of photometric redshifts
based on five and seven passband data.  The final redshift
distribution for all quasar candidates identified in the present work,
including those with poor classification, is shown in
Figure~\ref{fig:zfinal}, using the optical and, whenever possible, the
optical/infrared data.

\begin{figure}[ht]
\centerline{
\psfig{figure=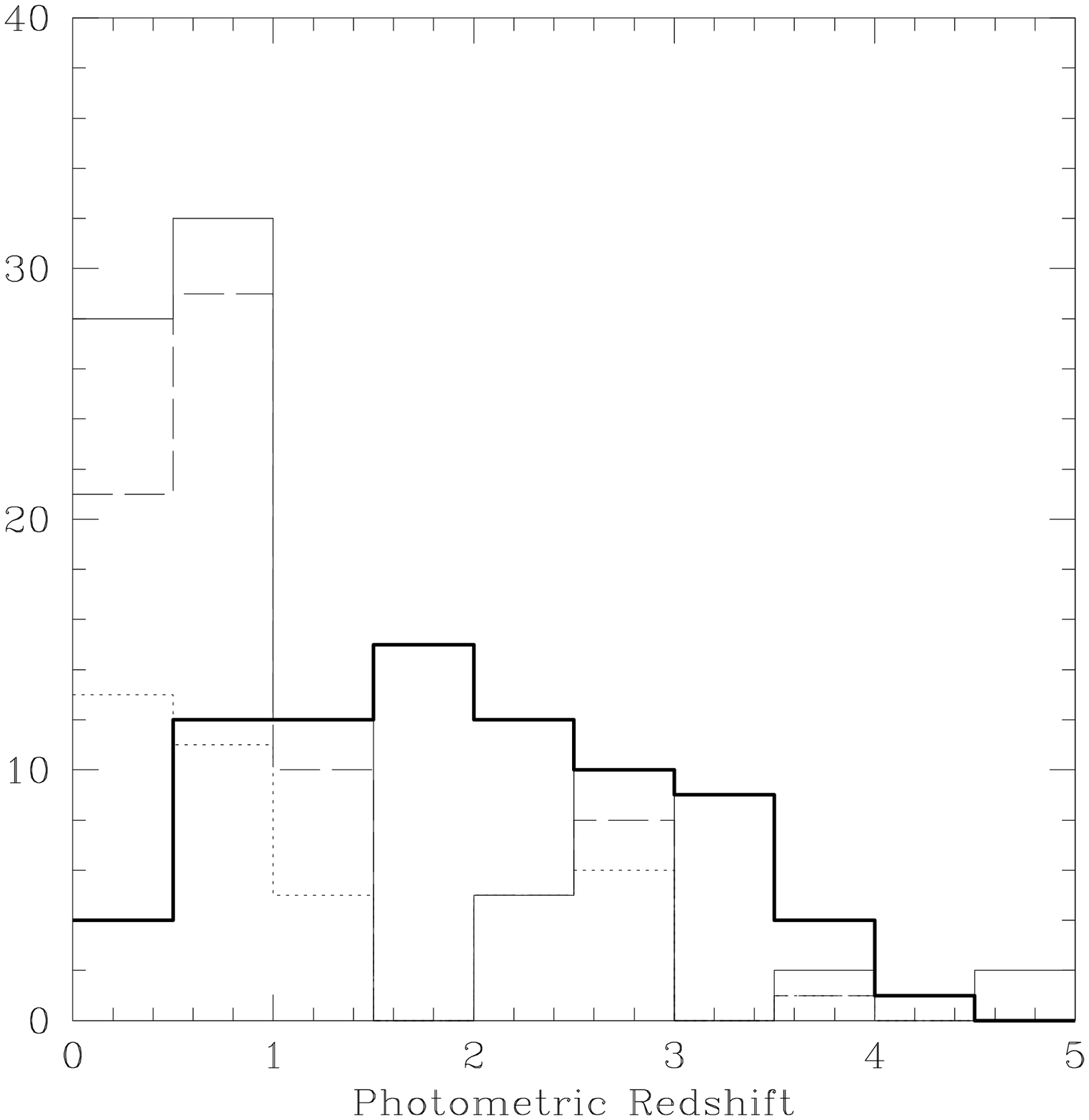,width=8cm,height=8cm}}
\caption{Photometric redshift distribution of quasar candidates
$\chi^2$-selected in the area covered by the optical and infrared
data. The figure shows the model predicted redshift distribution
(thick solid line) and the one measured for objects with no
$\chi^2$ selection (solid line), at 99\% (dotted line) and at 95\%
(dashed line)}
\label{fig:zphot7}
\end{figure}

\begin{figure}[ht]
\centerline{
\psfig{figure=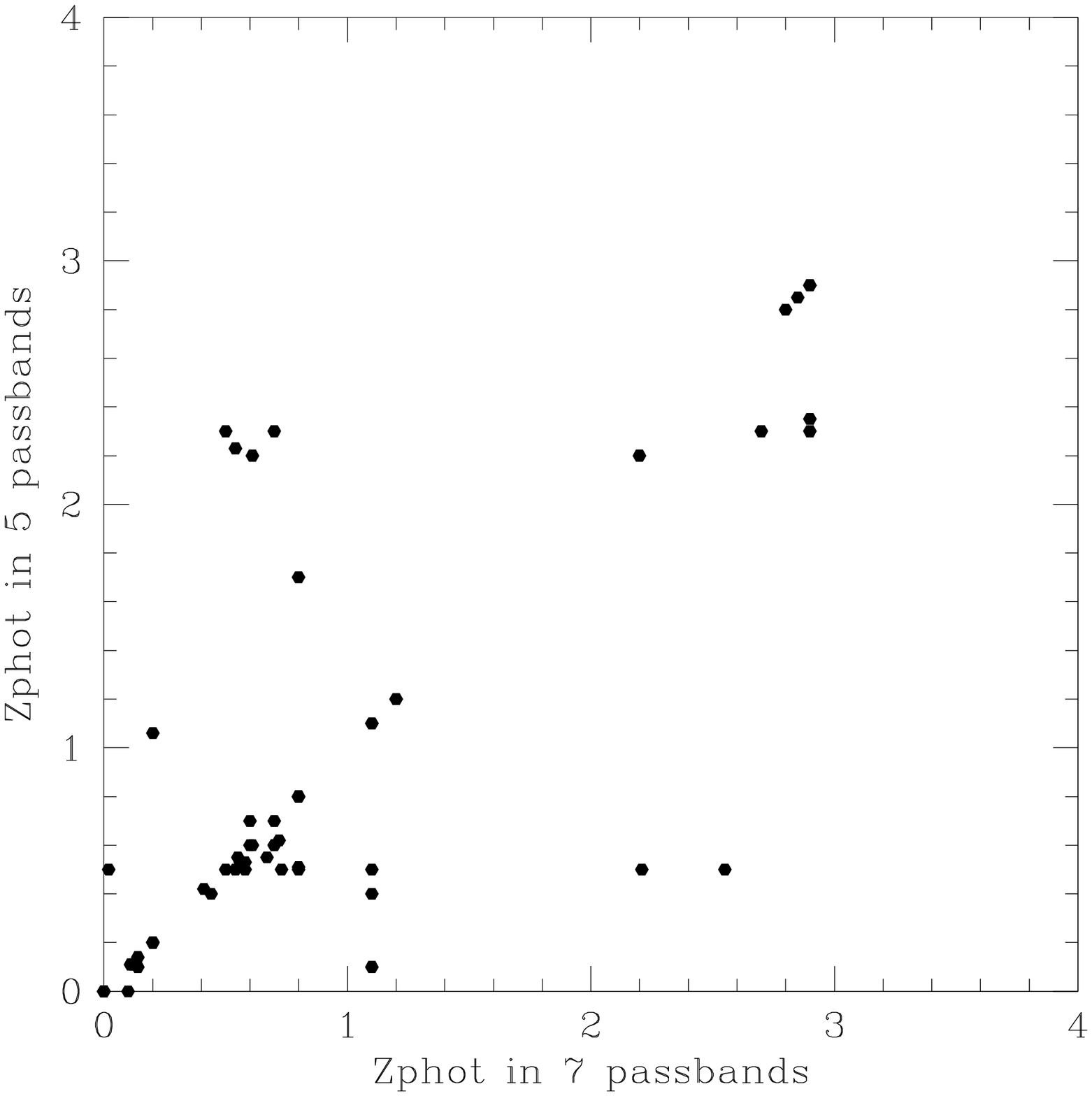,width=8cm,height=8cm}}
\caption{Comparison of the photometric redshift distribution of the 63
common quasar candidates using the five and seven passbands.} 
\label{fig:z7vsz5}
\end{figure}

Using the above redshift distribution one finds a surface density of
$\sim$55 quasar candidates with redshifts in the range $2.5\lsim
z\lsim3.5$ per square degree, within the area covered by the optical
observations. By contrast, using the optical/infrared data one finds a
surface density of $\sim$100 per square degree, thereby improving the
completeness of the sample.

\begin{figure}
\centerline{
\psfig{figure=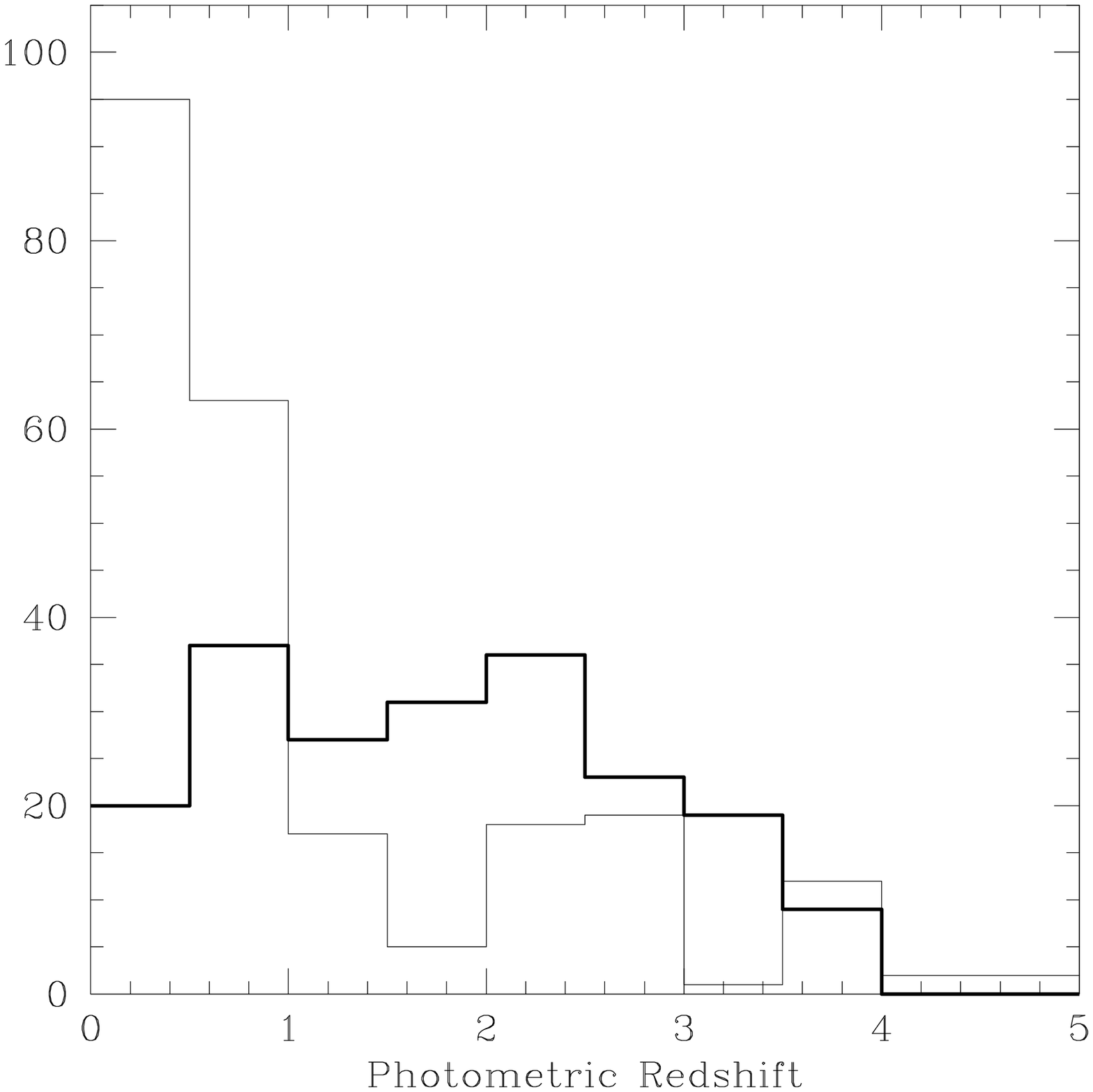,width=8cm,height=8cm}}
\caption{Final photometric redshift distribution for all 234
quasar candidates. The thick solid line shows the model predictions.}
\label{fig:zfinal}
\end{figure}

Table~\ref{fig:tabqso} lists the first 40 entries of the final quasar
candidate sample comprising 234 objects, extracted from the CDF-S.
The table gives the following information: in column (1) the EIS
identification number; in columns (2) and (3) the J2000 coordinates;
in column (4) the $R$-band magnitude (Vega system); in columns (5) -
(10) the optical/infrared colours; in column (11) the estimated
photometric redshift; in column (12) the ranking (1 - 3) of the
candidate based on the \xi2 confidence level, with the most robust
classifications denoted by rank equal to one; and in column (13) notes
as described in Section~\ref{evaluation}. The complete table can be
retrieved from the URL
``http://www.eso.org/eis/eis\_rel/dps/dps\_rel.html''. Note that in
the complete table the four candidates identified in optical but
rejected when using the infrared data have been included.

\begin{table*}
\caption{First 40 entries of the CDF-S quasar candidate list. 
$R$-magnitudes are given in the Vega system.}
\centerline{
{\psfig{figure=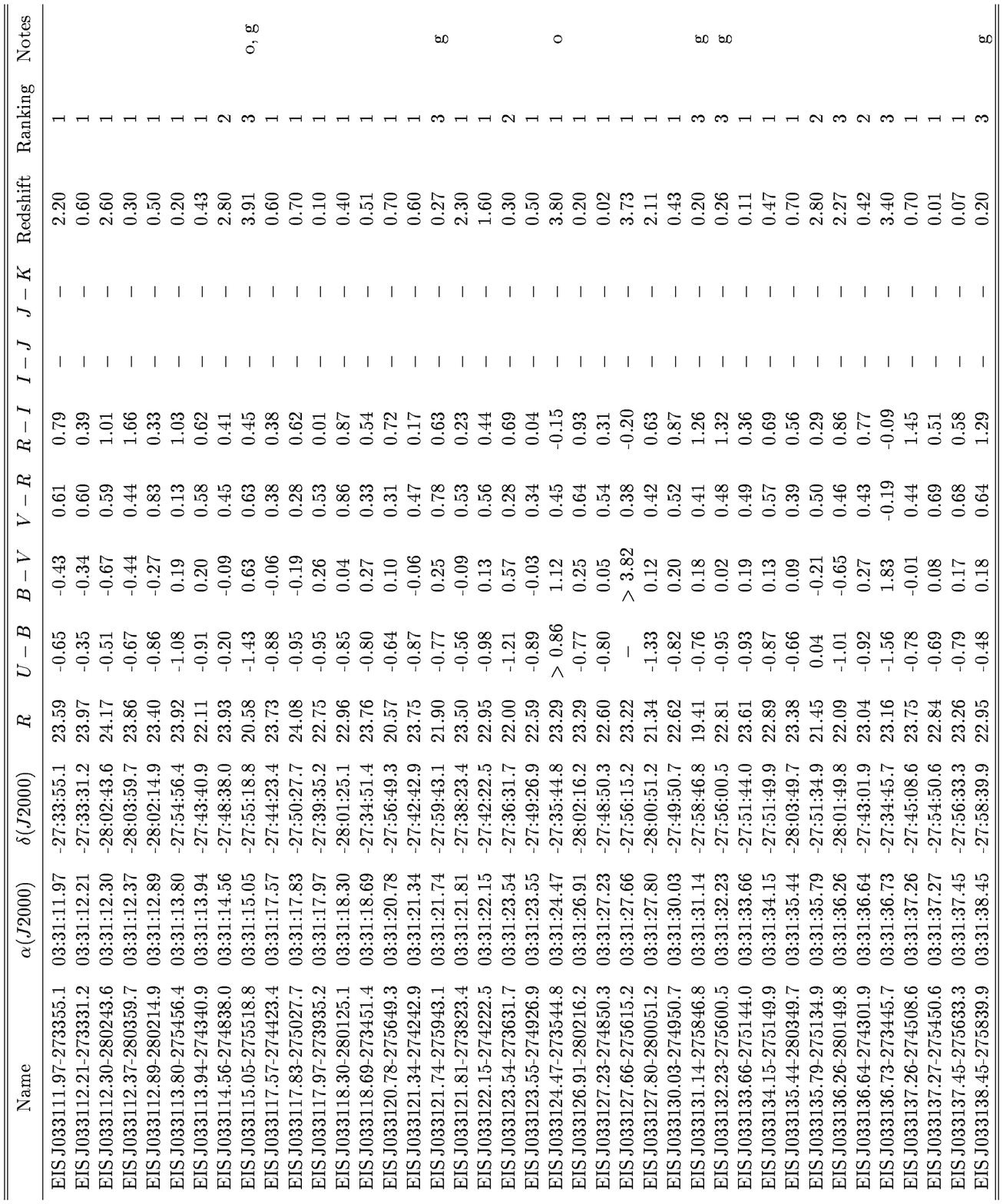,angle=0,width=\textwidth,clip=}}}
\label{fig:tabqso}
\end{table*}

\subsection {Galactic Objects}

The \xi2-technique primarily used for classification of quasar,
galaxies and stars has been extended to consider stellar sub-classes
such as white dwarves, low mass stars and brown dwarves. Even though
confirmation of these classifications will depend on spectroscopic
data this method is applied here as a first attempt to define a robust
procedure to select different sub-classes of galactic objects.

\subsubsection {White Dwarves}

In order to search for white dwarf candidates, 66 theoretical spectra
were provided from D. K\"oster, as well as three observed spectra of
very cool white dwarves ($T_{\rm eff} < 4000$K) from Ibata (2000;
F351-50, F821-07) and Oppenheimer (2001; WD0346).  The template
spectra were again compared to the broad-band photometry.  A total of
97 objects were classified as white dwarf candidates in the magnitude
range $20\lsim B\lsim25$, out of which 86 are robust
classifications. The number of 97 candidates is a factor of 1.5 higher
that the 65 white dwarves with $\log g > 7$, brighter than $B$=25,
expected to be found within the area of 0.25~square~degrees, as
predicted by current estimates of the white dwarf luminosity function
(Girardi et al., 2001).  From the total number of candidates, 89 have
estimated temperatures in the range from 6000K to 16000K according to
the theoretical spectra. The remaining nine were better matched by the
observed very cool white dwarf spectra, with seven being robust
classifications. The distribution of all candidates in the
$(U-B)/(B-V)$ plane is shown in Figure~\ref{fig:UBBVwd}. Two of the
cool white dwarves are located at $(B-V)\sim0.7$ and in the interval
$1.0<(U-B)<1.5$, superposing the main-sequence locus. The location of
these candidates are by and large consistent with the white dwarf
cooling curve kindly provided by P. Bergeron (2001). It is interesting to
point out that one of the tracks, representing the cooling curve of
hydrogen white dwarves curves towards the main-sequence. Note that
five of the cool candidates are U-dropouts and do not appear in this
colour diagram.

Comparison of Figure~\ref{fig:UBBVwd} with Figure~\ref{fig:UBBVqso}
shows that for the characteristics of the present survey a simple UVX
selection would lead to a large contamination of this sample by quasar
candidates, since these populations have a significant overlap in this
diagram. In fact, choosing typical values for the colours to delineate
the region occupied by white dwarves in this diagram, the fraction of
white dwarf candidates would correspond to about 30\% of the total
number of objects within this region. This is in contrast with the
high success rate (72\% spectroscopically confirmed) obtained by
Christlieb et al. (2001) in their analysis of the Hamburg/ESO Survey
(HES). These results illustrate the strong dependence of the
efficiency of colour selection with the characteristics of the survey.
A bright survey like HES would yield a low quasar and a high white
dwarf surface density while exactly the opposite is true for the deep
observations considered here.

\begin{figure}[ht]
\centerline{
\psfig{figure=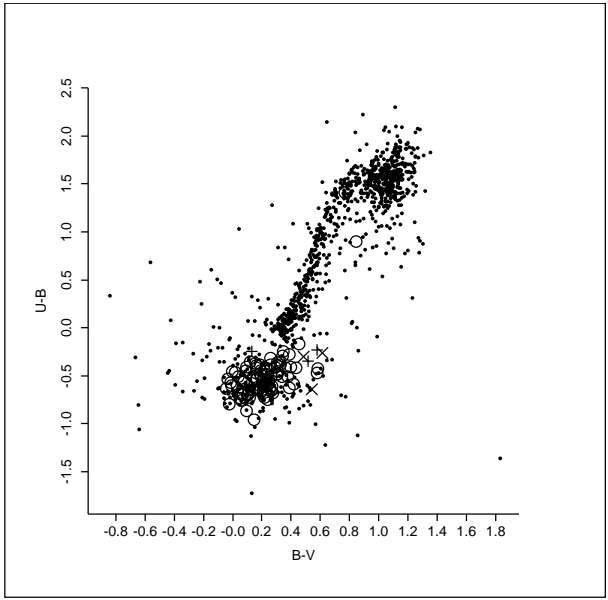,width=8cm}}
\caption{$(U-B)/(B-V)$ colour-colour diagram showing the \xi2-selected
white dwarf candidates within the area covered by the optical data.}
\label{fig:UBBVwd}
\end{figure}

\begin{figure}[ht]
\centerline{
\psfig{figure=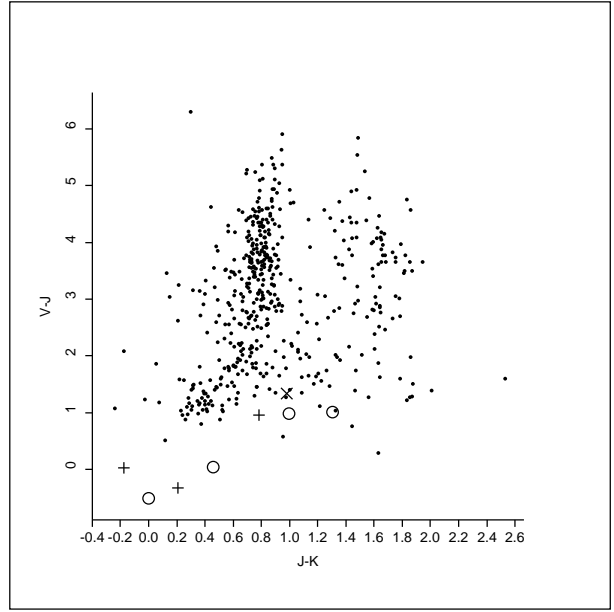,width=8cm}}
\caption{$(V-J)/(J-K)$ colour-colour diagram showing the \xi2-selected
white dwarf candidates within the area covered by the optical/infrared data.}
\label{fig:VJJKwd}
\end{figure}

The results obtained from the \xi2-analysis of the optical/infrared
data are as follow: a total of 21 candidates are selected with 18
being robust detections.  Figure~\ref{fig:VJJKwd} shows the
distribution of the candidates in the same colour-colour diagram as
Figure~\ref{fig:kxqso}. The locus of the white dwarves in this diagram
is shifted redwards relative to that computed by P. Bergeron (2001).
The candidates have estimated effective temperatures in the range 6000
to 14000K.  The overlap between the sub-samples extracted from the
five and seven passband comprises 18 objects.  The remaining three
objects were originally classified as quasar candidates. Another 17
objects selected as white dwarf candidates based on the optical only,
are now classified as quasar candidates.  These results show how
difficult it is to distinguish between quasars and white dwarves, and
how useful the infrared data can be for that purpose.

It is worth pointing out that none of the cool white dwarf candidates
identified using the optical colours are confirmed when the infrared
colours are included in the analysis. This may be due to inadequacies
in the near-infrared part of the model spectra, which could also
explain the shift of the locus of white dwarf candidates mentioned
above. This point will be further investigated when more infrared
spectra become available.

Table~\ref{fig:tabwd} lists the first 40 entries of the white dwarf
candidate sample, comprising 100 objects. The format is the same as in
Table~\ref{fig:tabqso}, with the effective temperature for the best-fit 
model presented in column (11).

\begin{table*}
\caption{First 40 entries of the CDF-S white dwarf candidate list.}
\centerline{
\psfig{figure=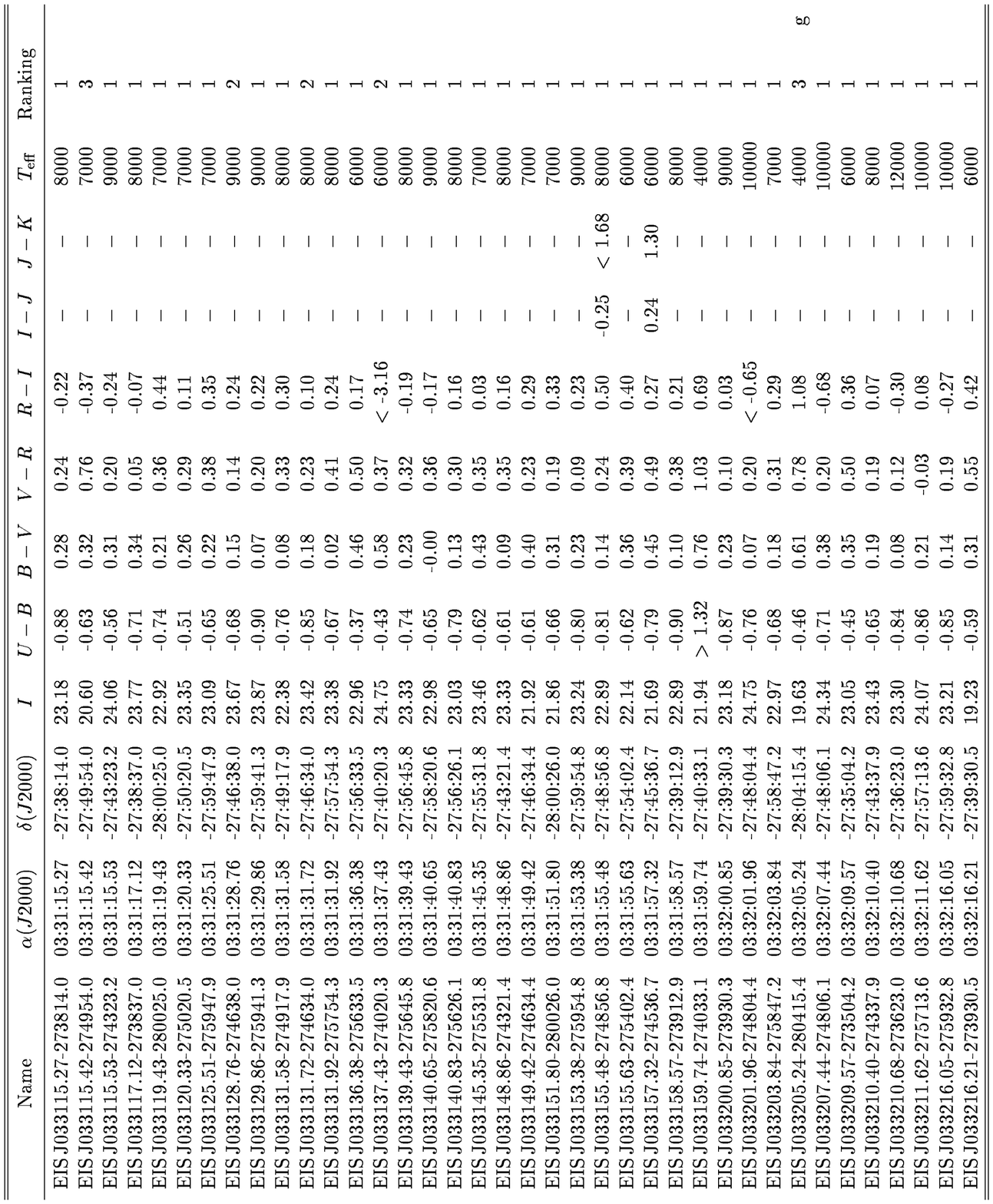,width=\textwidth,angle=0,clip=}}
\label{fig:tabwd}
\end{table*}

\subsubsection {Low Mass Stars and Brown Dwarfs}
\label{brown}

The low mass and brown dwarf spectral library was provided by Chabrier
and Baraffe and consists of 105 theoretical spectra. They correspond
to three sets of models which attempt to account for differences in
the formation and settling of dust in the atmospheres (Chabrier et
al., 2000). In this paper 53 of these models are used, corresponding
to objects with masses $\lsim0.1 M_{\odot}$ ($T_{\rm eff} \lsim
2800$K), to select low-mass stars and/or brown dwarf candidates. These
template spectra were compared to our broad-band SED and over the
0.25~square~degree area covered in five passbands a total of 18
candidates were identified with $18\lsim I\lsim22$ (all fainter than
$B\sim24$), with 13 being robust detections.  All of the candidates
are matched to spectra with effective temperatures between
1700K and 2800K, corresponding to masses roughly between 0.05 and 0.1
M$_{\odot}$, close to the hydrogen-burning limit. Their position on a
$(V-R)/(R-I)$ diagram is shown in Figure~\ref{fig:VRRIbd}. The objects
with $(R-I) > 2.5$ seen on this figure that have not been selected as
low mass or brown dwarf candidates were identified as M6V stars.  Note
that this class marks the transition between main sequence stars and
low mass stars. Comparing the results of the
\xi2-analysis with those obtained selecting objects
redder than $(R-I)>2.3$, roughly corresponding to the $(V-I)>3.5$
criterion adopted by Zaggia et al. (1999), one finds a significant
contamination ($\sim$75\%) by other types of objects.

Applying the $\chi^2$-test on the optical/infrared data one finds a
total of 35 candidates out of which 14 are robust classifications.
All five low-mass star candidates that have both optical and
near-infrared data are confirmed when the $J$ and $K_s$ information is
included in the analysis. Furthermore, 11 stars originally classified
as M5V and M6V stars using the $UBVRI$ catalogues, are classified as
low-mass stars when the infrared data are used.  All candidates lie in
the range $15.5\lsim K_s\lsim19.0$ and have estimated effective
temperatures in the range between 1700K and 2800K.  The candidates
identified are shown in Figure~\ref{fig:IJJKbd}. The figure shows two
main concentrations of low mass star candidates. One at
$(J-K_s)\sim0.8$ and $(I-J)\sim1.5$, with 19 candidates, roughly
corresponding to the transition between main sequence and very low
mass stars. The other covers the region defined by $(J-K_s)\gsim1.4$
and $(I-J)\gsim1.8$ (13 objects), consistent with the location of the
L-dwarves (L3) as reported in the literature (\eg Reid et al., 2001;
Leggett et al., 2001; Schweitzer et al., 2001). In this
optical/infrared colour diagram, as mentioned in the previous
sections, one sees again a population of objects with colours which
are not predicted by any model describing the spectral properties of
point-sources. As discussed below, most of these objects are
associated with unresolved galaxies which contaminate the point-source
catalogue.

\begin{figure}[ht]
\centerline{
\psfig{figure=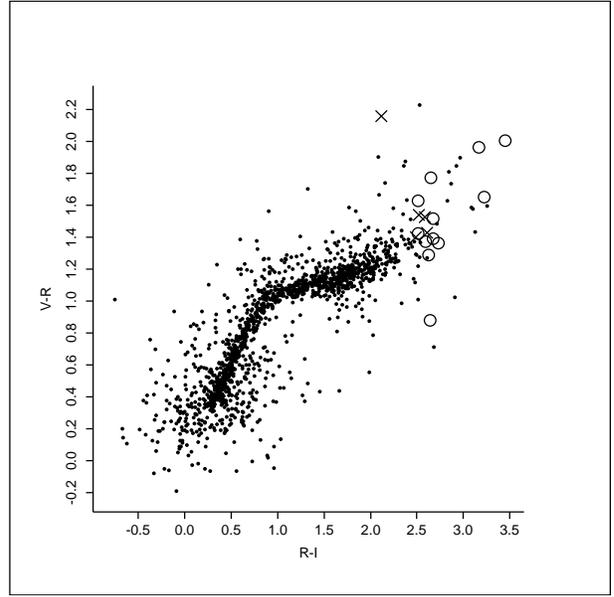,width=8cm}}
\caption{Optical colour-colour diagram showing the brown dwarf candidates
selected from the \xi2-technique.}
\label{fig:VRRIbd}
\end{figure}

\begin{figure}[ht]
\centerline{
\psfig{figure=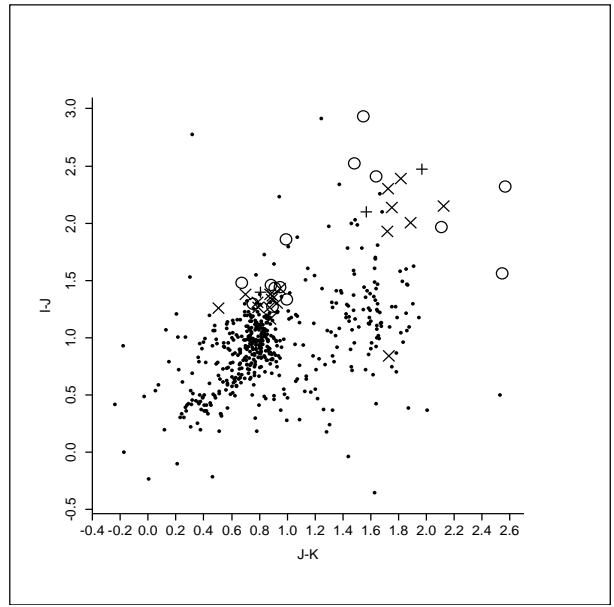,width=8cm,height=8cm}}
\caption{Optical/infrared colour-colour diagram showing the brown dwarf 
candidates selected using the \xi2-technique.}
\label{fig:IJJKbd}
\end{figure}

Based on the results of the \xi2-selection one finds a surface density
of very low mass stars of about 72 candidates per square~degree using
optical colours. When the near-infrared data is added, this value
increases by a factor of 3, yielding a surface density of 350 per
square degree. These estimates for the surface density are a factor of
3 higher than the expected value of 116 low-mass stars per
square~degree with $T_{\rm eff}\lsim2800$K and brighter than $B=25$,
predicted by models (\eg Girardi et al., 2001). 

The final list of individual low-mass star candidates in the CDF-S
field is given in Table~\ref{fig:tablms}. The table format is the same
as that of Table~\ref{fig:tabwd}.

\begin{table*}
\caption{List of low-mass star candidates in the CDF-S.}
\centerline{
\psfig{figure=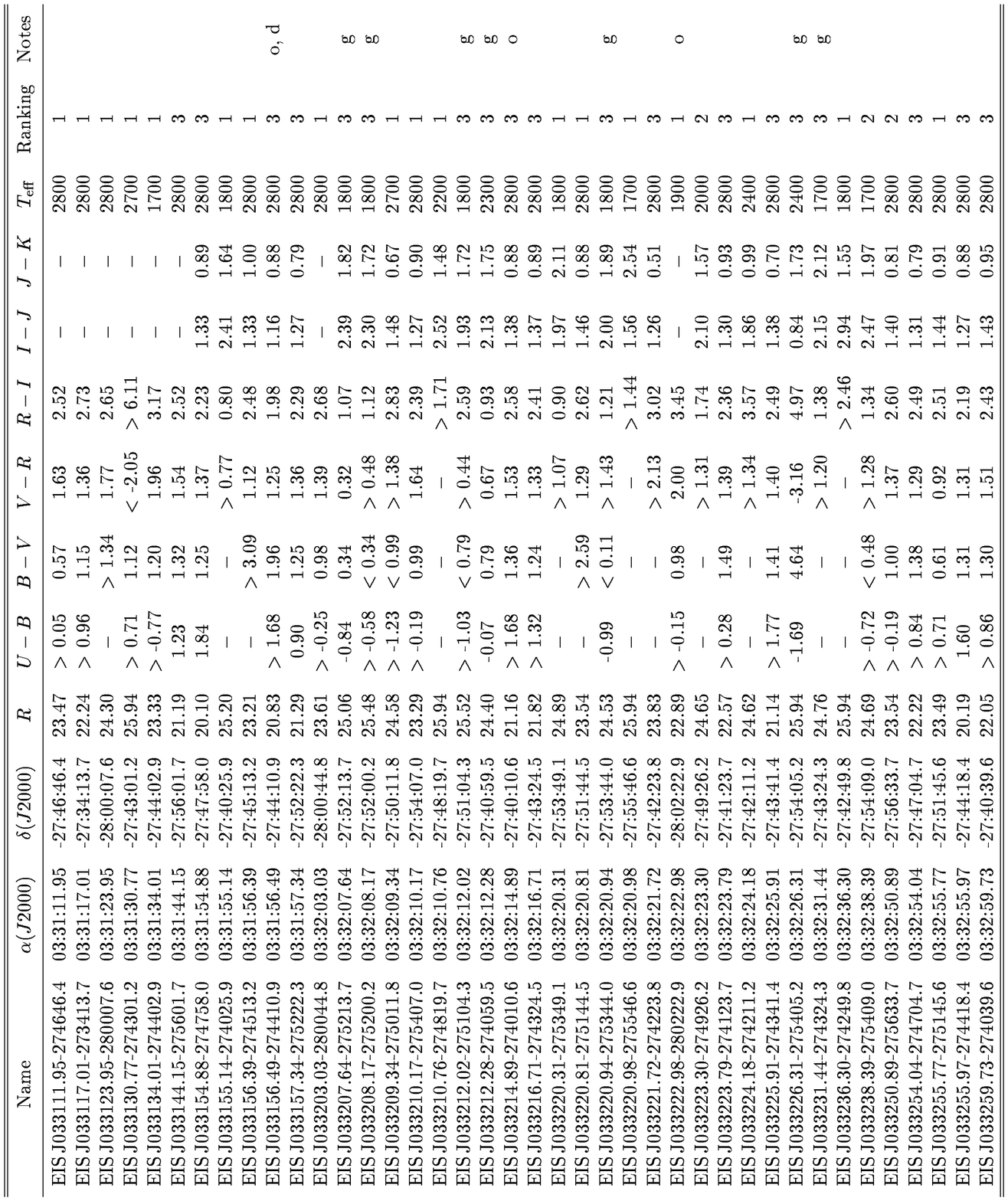,angle=0,width=\textwidth,clip=}}
\label{fig:tablms}
\end{table*}

\subsection{Very red objects}
\label{extreme}

In the previous sections only objects detected in at least three
passbands have been considered. There are, however, objects detected
in one or two passbands, that may be of interest as well and for which
neither the $\chi^2$-method nor colour-colour selection can be of
use. Of particular interest are those detected in the red-most
passbands available, \eg $R$ and/or $I$ over 0.25~square~degrees and
$J$ and/or $K_s$ for 0.1~square~degrees. While impossible to classify
them having a single colour and, in some cases, just a lower limit,
these very red objects are natural candidates for follow-up
spectroscopy. However, they could also be associated with possible
features in the colour catalogue production, as discussed in
Section~\ref{evaluation}. Table~\ref{tab:ero} lists the
identification, coordinates, colour (whenever available) and red-most
magnitude of the objects detected in $R$ and $I$ (16 objects), in $I$
only (4 objects), in $J$ and $K_s$ (5 objects), and in $K_s$ only (4
objects).

\begin{table*}
\begin{center}
\caption{Very red objects detected in the red-most filters available.}
\label{tab:ero}
\begin{tabular}{lccllc} 
\hline
\hline
 {\rm Name}        & $\alpha (J2000)$ & $\delta (J2000)$ & $I$ & $R-I$ & {\rm Notes}\\
\hline
\hline
\\
EIS\;J033112.31-275640.1 & 03:31:12.31 & -27:56:40.1 & 21.86  & 2.53    \\
EIS\;J033119.40-274834.9 & 03:31:19.40 & -27:48:34.9 & 21.73  & 3.75    \\
EIS\;J033140.38-275942.9 & 03:31:40.38 & -27:59:42.9 & 20.50  & 1.79  &  d \\
EIS\;J033141.54-273505.2 & 03:31:41.54 & -27:35:05.2 & 21.64  & 2.94    \\
EIS\;J033143.33-274707.3 & 03:31:43.33 & -27:47:07.3 & 21.71  & 2.79    \\
EIS\;J033155.88-280344.7 & 03:31:55.88 & -28:03:44.7 & 21.22  & 2.58    \\
EIS\;J033221.01-273659.1 & 03:32:21.01 & -27:36:59.1 & 20.89  & 2.12  &  d \\
EIS\;J033226.53-273721.2 & 03:32:26.53 & -27:37:21.2 & 21.60  & 3.38    \\
EIS\;J033226.64-280248.9 & 03:32:26.64 & -28:02:48.9 & 20.71  & 3.76    \\
EIS\;J033249.39-273405.8 & 03:32:49.39 & -27:34:05.8 & 21.45  & 3.75    \\
EIS\;J033254.87-280224.1 & 03:32:54.87 & -28:02:24.1 & 21.47  & 3.99    \\
EIS\;J033257.33-280310.1 & 03:32:57.33 & -28:03:10.1 & 22.00  & 2.68    \\
EIS\;J033302.20-275914.4 & 03:33:02.20 & -27:59:14.4 & 21.80  & 3.29    \\
EIS\;J033317.64-274040.0 & 03:33:17.64 & -27:40:40.0 & 21.94  & 1.58  &  d \\
EIS\;J033327.41-280253.5 & 03:33:27.41 & -28:02:53.5 & 20.95  & 4.07    \\
EIS\;J033342.70-274618.1 & 03:33:42.70 & -27:46:18.1 & 21.87  & 1.91    \\
EIS\;J033132.87-274111.4 & 03:31:32.87 & -27:41:11.4 & 21.11&$>$4.9 & d \\
EIS\;J033211.00-275904.7 & 03:32:11.00 & -27:59:04.7 & 21.86&$>$4.1 & d \\
EIS\;J033230.20-273337.6 & 03:32:30.20 & -27:33:37.6 & 21.12&$>$4.8 & d \\
EIS\;J033251.60-275917.5 & 03:32:51.60 & -27:59:17.5 & 21.57&$>$4.4 & d \\
\\
\hline
\hline
{\rm Name}  & $\alpha (J2000)$ & $\delta (J2000)$  & $K_s$ & $J-K_s$ & {\rm Name}\\
\hline
\hline
\\
EIS\;J033229.58-274812.5 & 03:32:29.58 & -27:48:12.5 & 19.86 & 2.24   & \\
EIS\;J033238.14-274750.1 & 03:32:38.14 & -27:47:50.1 & 20.33 & 1.37   & \\
EIS\;J033241.92-274512.5 & 03:32:41.92 & -27:45:12.5 & 20.32 & 1.29   & \\
EIS\;J033245.80-274211.4 & 03:32:45.80 & -27:42:11.4 & 19.98 & 1.13   & d \\
EIS\;J033304.21-275137.3 & 03:33:04.21 & -27:51:37.3 & 19.25 & 1.00   & d \\
EIS\;J033225.12-274219.7 & 03:32:25.12 & -27:42:19.7 & 19.09 & $>$4.3 & d \\
EIS\;J033226.12-274327.0 & 03:32:26.12 & -27:43:27.0 & 20.42 & $>$3.0 & \\
EIS\;J033242.43-274236.7 & 03:32:42.43 & -27:42:36.7 & 19.71 & $>$3.7 & d \\
EIS\;J033307.51-274435.6 & 03:33:07.51 & -27:44:35.6 & 20.18 & $>$3.2 & d \\
\hline
\hline
\end{tabular}
\end{center}
\end{table*}

Figure~\ref{fig:red} shows the $(R-I) \times I$ colour-magnitude
diagram for all point-sources within the area of 0.25~square~degrees
(left panel) and the $(J-K_s) \times K_s$ diagram for the central area
of the CDF-S covered by infrared data (right panel). The symbols are
described in the figure caption. The extreme colour lower limits of
the $R$- and $J$-dropouts shown in the figure make them likely to be
brown dwarves (but see Section~\ref{evaluation}).

\begin{figure*}
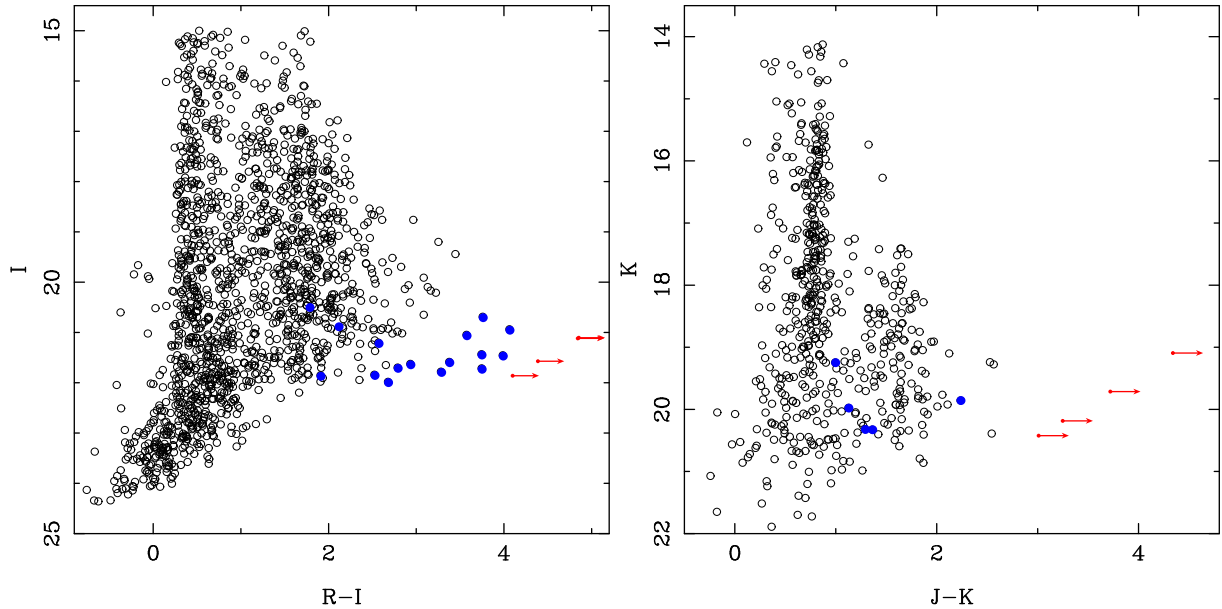

\centerline{
\psfig{figure=IRI.ps,height=8cm,width=8cm}
\psfig{figure=KJK.ps,height=8cm,width=8cm}}
\caption{Colour - magnitude diagrams for the very red objects:
$I$ versus $R-I$ (left panel) and $K_s$ versus $J-K_s$ (right
panel). The filled circles show objects detected in two bands. Arrows
indicate lower limits in the colour of objects detected only in the
red-most passband.}
\label{fig:red}
\end{figure*}

\subsection{Outliers}

As discussed in Section~\ref{searchout} there are several reasons why
one would like to search for outliers. From a pure technical point of
view, objects with odd colours have to be identified and visually
inspected as they may reveal problems in the construction of the
colour catalogue, contamination by close neighbours, cosmic rays or
other image artifacts. Alternatively, they may represent potentially
interesting rare cases, either of known objects such as quasars at
very high-redshifts or previously unknown populations. Therefore,
classifying objects as outliers is an important step towards verifying
the integrity of the colour catalogue and avoiding overlooking new
discoveries.

\begin{figure}
\centerline{
\psfig{figure=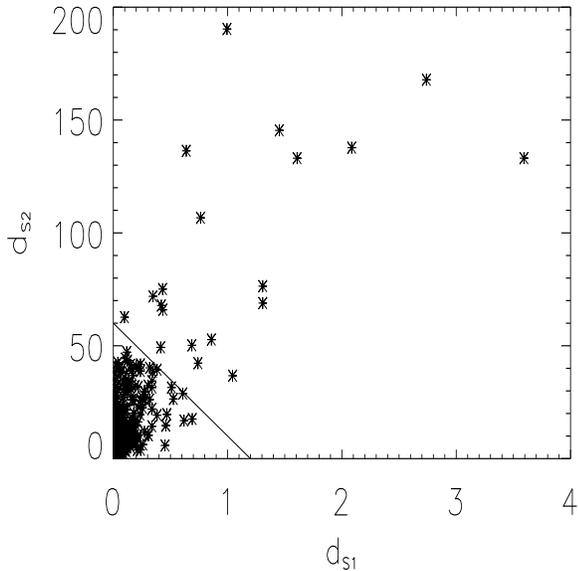,width=9cm,height=9cm}}
\caption{Illustration of the selection of outliers applied on the
five passband sub-sample for $m=2$. For definitions of d$_{S_1}$ and
d$_{S_2}$ see Section~\ref{methods}.}
\label{fig:outliers}
\end{figure} 

As described in Section~\ref{searchout}, outliers are identified in
colour space based on their distances d$_{S_1}$ and d$_{S_2}$ from
their nearest neighbour. An isolation criterion is then chosen,
depending on the position of the objects on the d$_{S_1}$ versus
d$_{S_2}$ diagram, as schematically shown in
Figure~\ref{fig:outliers}. This criterion divides the d$_{S_1}$ versus
d$_{S_2}$ space in two regions, a densely populated one towards low
values of the parameters and a much less dense region, where outliers
lie. The parameters describing the separation line are chosen by
fine-tuning them to include the most obvious cases of isolated objects
in different colour-colour projections, separately for $m=2$ and
$m=3$. Note, however, that objects isolated in one of the
colour-colour projections are not necessarily isolated in all of them.

\begin{table}
\caption{Number of objects identified as outliers for different samples.}
\label{tab:out}
\begin{tabular}{lccc}
\hline
\hline
Sub-sample & N & $m=2$ &  $m=3$  \\
\hline
\hline
\\
$UBVRI$ & 1164 & 19 & 26 \\
$BVRI$ & 300 & 13 & 17 \\
$UBVRIJK_s$ & 385 & 19 & 18 \\
$BVRIJK_s$ & 119 & 13 & 19 \\
\hline
\hline
\end{tabular}
\end{table}

The results from the outlier analysis are summarised in
Table~\ref{tab:out}, which lists the sub-samples, the number of
objects in them and the number of outliers for $m=2$ and $m=3$,
respectively. Note that in general the number of outliers increases
with $m$.  From the table one finds that the fraction of outliers is
small being typically $\sim$10\% of the whole sample. For both $m=2$
and $m=3$, about 60\% of the outliers are indeed poorly classified by
the \xi2-method, while 30\% are robust candidates. These results
indicate, as expected, that the outliers consist of a mix population
including known rare objects, objects possibly not well described by
the available spectral library, or undesirable features on the images
or the derived catalogues.  The total number of outliers is 64 (80)
for the sub-samples analysed using $m=2$ ($m=3$), out of which about
60\% deserve a closer investigation as presented in the next Section.

Figure~\ref{fig:out} illustrates the location of outliers identified
by applying the methodology described above to the five (seven)
passband sub-sample. The figure shows two projections of the colour
space, one for each of the sub-samples considered. The different
symbols represent outliers selected using different values of
$m$. Since nearly all of the outliers identified using $m=2$ are also
identified when $m=3$ is used, in these plots only the additional
objects identified with $m=3$ are represented by a different symbol.
While a single projection is not sufficient to determine whether an
object is truly an outlier in the multi-dimensional colour space, most
objects far from the main concentration of points are successfully
identified. In particular, in the left panel one finds the object with
very particular colours, mentioned in section~\ref{quasar}, originally
classified as a quasar. This case as well as others will be discussed
in the next Section.

Generally speaking, among the 30\% of outliers which are also robust
\xi2 classifications about half are identified as quasar candidates and
half are identified as galactic object candidates from the \xi2-technique. In
both cases, the candidates found to be outliers are associated with
sparse populations with nearly all quasars having $z\gsim3$ and most
of the stars being early spectral types (O-A) which are rare, especially at
high-galactic latitudes. These results apply equally well to all the
sub-samples and isolation criteria adopted.

Note that the selection of the $m-th$ neighbour as well as the
separation line must be empirically determined. Even thought there is
a correspondence between the outlier selection and the values of the
$\chi^2$, the exact relation is not easy to establish.

\begin{figure*}
\centerline{
\psfig{figure=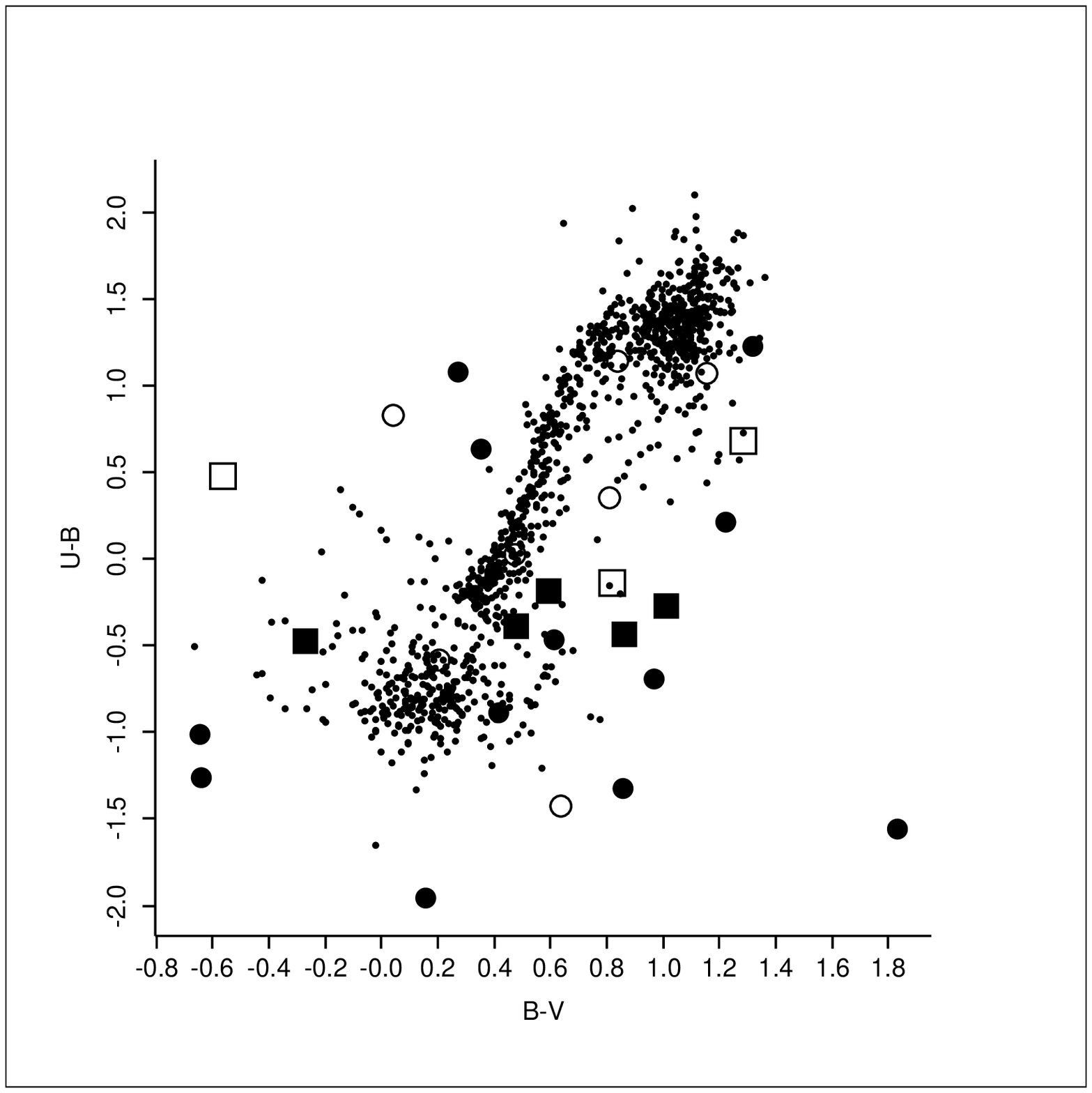,width=8cm,height=8cm}
\psfig{figure=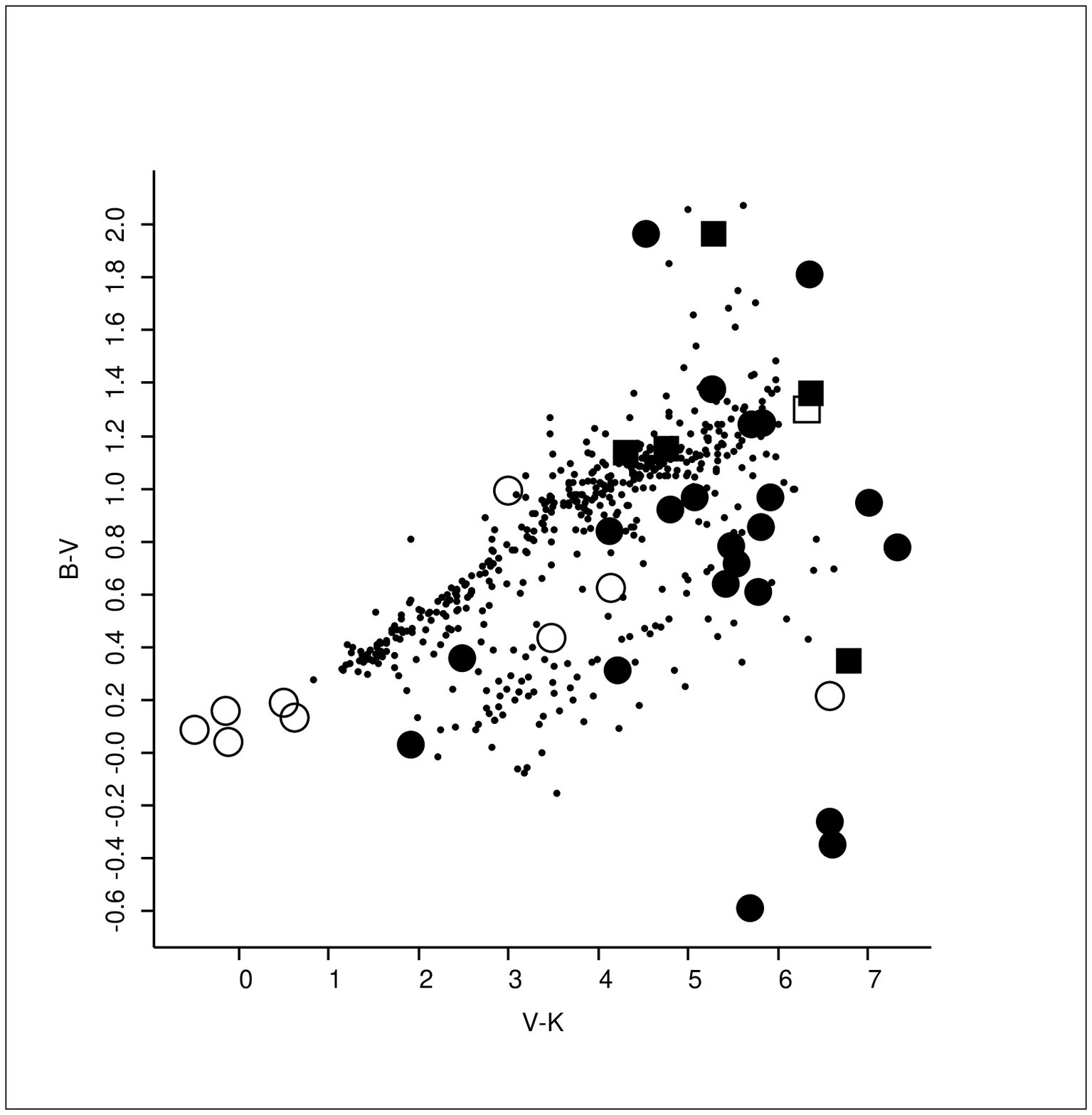,width=8cm,height=8cm}}
\caption{$(U-B)/(B-V)$ (left panel) and $(B-V)/(V-K_s)$ (right panel)
 showing objects classified as outliers by the criteria given in
Section~\ref{methods} adopting $m=2$ (circles) and $m=3$
(rectangles). Open (filled) symbols denote objects with good (poor)
classifications.}
\label{fig:out}
\end{figure*}

\section{Evaluation of Classification Results}
\label{evaluation}

In previous sections the results of the \xi2 classification were
tentatively assessed by comparing these classifications with those
that would be obtained from selecting regions in two-colour
diagrams. In this section this evaluation is complemented by an
investigation of the nature of the outliers and extreme red objects
detected in two or only one band, in several cases resorting to a
visual inspection of the images and a careful examination of the
photometric measurements obtained for these objects. The motivation is
to find possible effects that may impact the \xi2 fitting and lead to
a poor classification. The primary goal is to investigate whether
these cases are of interest or are, instead, associated with
particularities of the images and/or derived source catalogues for
which, perhaps, additional refinements in the selection criterion can
be devised. The measured SED of the poorly classified objects are also
compared to a set of galaxy models to evaluate the possible
contamination of the point-source catalogue by unresolved galaxies.
The information stemming from the discussion below is incorporated as
notes to the
Tables~\ref{fig:tabqso},~\ref{fig:tabwd},~\ref{fig:tablms}, and
\ref{tab:ero}.  In these notes ``o'' denotes outliers, ``d'' objects
that should be discarded from further consideration, and ``g'' objects
which are better matched by galaxy spectra.

\subsection {Nature of Outliers}

Out of a total of 81 outliers, representing less than 5\% of the total
sample of point-sources, image cutouts were produced for all 57
objects having poor classification based on the
\xi2-analysis. From the visual inspection of their images the following
conclusions can be drawn. About 44\% of the cases (25 objects) are
found to be significant SExtractor detections, point-like sources and
isolated. This set of objects consists of a mixed population which
includes predominantly and high-redshift quasars ($2.1 <z < 3.94$)
candidates.  Among them there are two interesting cases, both poorly
classified as quasars, one of which has already been mentioned in
Section~\ref{quasar} (Figure~\ref{fig:UBBVqso}).  Close examination of
their photometry shows that both objects have comparable magnitudes in
all passbands except in $B$ in which they are more than one magnitude
fainter. Inspection of the original images show no reason to suspect
problems in the magnitude estimates. The only possible explanation for
the odd colours is that these objects may be variable. This seems to
be a reasonable explanation considering that images taken in all
passbands except $B$ were obtained a few days apart, while the
$B$-images were taken nine months earlier. This result points out the
importance of including temporal information, whenever possible,
in the type of analysis presented here. 

About 25\% of poorly classified outliers are located near
other objects leading to problems of contamination, de-blending and
erroneous associations and should be discarded.  The remaining 30\% of
the cases have been identified with features of the
source extraction algorithm and with the production of colour
catalogues, which are now being addressed.

In summary, about 25\% of the outliers are robust
classifications. Another 25\% are isolated objects with no apparent
problems in their photometric measurements, indicating that their poor
classification may be due to inadequacies in the spectra
library. These objects are thus prime targets for spectroscopic
observations.  These cases as well as outliers found to be potentially
problematic are indicated in the tables of different classification
types presented in previous sections.  Taken the above numbers at face
value one can conclude that only for about 2\% of the objects in the
original point-source colour catalogue should be discarded because of
problems, most of which unavoidable in nature being caused by the
presence of nearby objects affecting their photometric measurements.

\subsection{Very Red Objects}

In addition to the outliers one should also carefully examine extreme
cases such as those presented in Section~\ref{extreme}.  From the
total of 29 objects listed in Table~\ref{tab:ero}, 17 showed no
problems when visually inspected. Out of the remaining 12, seven are
close (less than 4 arcsec) to brighter objects or located on the outer
parts of a galaxy, and were not properly de-blended.
Included are two $K$-only objects listed in Table~\ref{tab:ero}.
Another related case is one object found close to one of the masks
automatically placed around a very bright star. As before, the
photometric measurements are not reliable.  Two additional problems
were recognized. One is the break-up of a galaxy image by the source
extraction algorithm (1 case) and the detection of residual cosmic
rays in regions poorly sampled by the dithered images (2 cases, both
$I$-only detections).  Finally, one $JK_s$-only object, while visible in
the $I$-band image, was not detected.

\subsection {Contamination by Galaxies}

While the bright catalogue considered here should, in principle,
contain only point sources, objects having poor classification (see
Section~\ref{methods}) according to the \xi2-method have also been
compared to galaxy templates. Overall there are 216 and 221 objects
with poor classification in the five and seven passband catalogues,
respectively.  From their comparison with galaxy spectra one finds
that 25 - 35\% have an improved fit, with 51 and 71 objects being
classified as galaxies, respectively.  The objects classified as
galaxies have mix morphological classifications and cover a broad
redshift range. Based on these numbers one estimates that a
lower-limit for the contamination of the five and seven passband
point-source catalogues by unresolved galaxies is of the order of 5 -
10\%.


\section{Discussion}
\label{discussion}

The methodology described in this paper has been developed to analyse
in an objective and automatic way colour catalogues being routinely
produced by the EIS pipeline in order to: assign objects to different
classes of astronomical sources; allow for new discoveries; and
understand the limitations of the data and the procedures adopted in
the derivation of source catalogues and their colours from the
association of data taken in different passbands. The ultimate goal is
to define procedures to efficiently extract from imaging survey data
well-defined samples, with minimal contamination, for spectroscopic
follow-ups.

As a first step in this direction the method of \xi2 fitting template
spectra to the measured broad-band photometry, currently being used
for estimating galaxy/quasar photometric redshifts, has been employed,
extending it to include different types of galactic objects. The
method is intended to replace standard classification schemes based on
the analysis of one or more two-colour diagrams which becomes
unmanageable for large sets of multi-band data. As currently
implemented the classification scheme only considers the spectral
properties of the objects, neglecting other important information as
the apparent magnitude of the objects and the expected density of
objects of different types (see below).  The results obtained from the
automatic classification are not only consistent with those that would
have been obtained from traditional methods based on two-colour
diagrams but also consistent with model predictions, while minimising
contamination by objects of other types.  A point worth noting is that
a significant number of poor classifications stem from the fact that
the passbands used in the present analysis not always provide
independent information and the statistical analysis leads to
artificially low significances of the resulting classifications. In
order to deal with this problem other techniques which take into
account the proper dimensionality of the colour space for a specific
class of objects (\eg PCA) should also be considered.  Finally, it is
worth emphasising that the classifications are as good as the
available spectral library. The library currently being used has been
assembled from publicly available models and data and a number of
classes are under-represented. Improvements in the classification
method will depend on the continuous upgrade of the available spectral
library. Currently, the library is being upgrade to include infrared
spectra of white dwarves and low-mass stars kindly provided by
S. Leggett. Adding spectra for different type of objects from other
ongoing spectroscopic surveys such as the SDSS  will also be of great
value in improving the current library.

In order to detect potential problems and not to overlook possible new
discoveries the \xi2-method has been complemented by a procedure of
identifying outliers using as dissimilarity measures Euclidean
distances in the multi-dimensional colour space and adopting a
nearest-neighbour isolation criterion. Despite its simplicity the
criterion adopted identified rare population of objects, objects with
odd colours which could be traced either to real physical effects such
as variability or to problems with their measured colour,
demonstrating its usefulness in greatly reducing the number of cases,
about 5\% of the entire sample, that require a more detailed
inspection. This number could be reduced even more by a further
screening of the sample. As alluded to in the previous section,
information about variability, if available, is of great use as it is
a criterion based on angular separation and magnitude differences
between a source and its nearest neighbour or to the mask
automatically placed around very bright stars. This together with
SExtractor de-blending flag and distance to masks placed around very
bright stars should be used to discard objects which are likely to
have the photometry affected by light contamination, the most common
problem identified. 

Overall, the present analysis suggests that derived catalogues are
mostly free of problems. Visual inspection of several odd colour or
extremely red objects has revealed that the most frequent problems are
associated to the limitations of the de-blending algorithm;
contamination by close neighbours; and, in some cases, residual cosmic
rays located in poorly sampled regions of the image mosaic, with
insufficient number of stacked images for a proper
sigma-clipping.

In a follow-up paper the classification based on the spectral
properties, as presented here, will be complemented with other
statistics which further characterize the different populations of
extragalactic/galactic objects using mock catalogues created from
Monte Carlo simulations. This is particular important in the analysis
of point-sources for which the stellar population makes an important
contribution which varies according to the position of the sky
observed. To account for this as well as reddening effects and the
different filter sets used by the various surveys a population
synthesis model has been combined with galactic structure models to
simulate different observations.

\section{Summary}
\label{summary}

This paper describes the methodology developed to analyse
multi-wavelength data from ongoing public surveys to objectively and
automatically classify extracted objects based on their colours. The
method is expected to yield samples with better completeness and less
contamination, than previous analysis based on two-colour
diagrams. Moreover, the analysis can be carried out automatically and
thus better cope with the rapidly increasing data volume from imaging
surveys and more efficiently producing improved samples (higher yield)
to feed large-aperture telescopes. The method has been applied to a
catalogue of point-sources extracted from the optical/infrared images
taken of the CDF-S field by the EIS project.  The CDF-S field is of
particular interest in this context considering the number of
spectroscopic and photometric programs either ongoing or planned for
the near future.  These data will be invaluable to directly assess the
results of the present analysis which may lead to refinements in the
classification method adopted. The paper also provides a rough
estimate of the by-products that should be expected at the end of the
ongoing DPS.

The data used consists of $UBVRI$ images taken with a wide-field
imager covering an area of 0.25~square~degree, complemented by a
mosaic of infrared $JK_s$ images covering
0.1~square~degrees. Combining these data one finds a total of 234
quasar candidates with estimated photometric redshifts up to $z\sim5$,
among which 16 have $z\gsim3.5$. In addition, 51 low-mass star/brown
dwarf candidates and 100 white dwarf candidates were identified.
Tables listing the classified objects are presented together with the
properties inferred by the classification method. If the
classifications presented here are confirmed, samples comprising over
200 high-redshift quasars ($z>$3.5), and over 1000 white dwarves,
$\sim$100 cooler than 4000K, will become available at the end
of the survey, expected to cover 3~square~degrees. 

It is worth emphasising the contribution of the near-infrared data. It
increases the accuracy of the determination of the photometric
redshifts as well as the number of quasar candidates in the redshift
interval $2.5\lsim z\lsim 3.5$. Infrared photometry is also essential
for tracking very low-mass stars and brown dwarves.  If only optical
data are considered, one should expect the DPS to provide $\sim$200
low-mass stars, including only $\sim$20 L-dwarves over the
3~square~degree area. Since the surface density of L-dwarves increases
to 120 per square degree when infrared data is considered, the number
of L-dwarves would be much larger if the infrared observations were
extended. As it currently stands the infrared observations cover an
area of 0.2 square degrees within which one expects $\sim 25$
L-dwarves.

The results presented here will be used in a forthcoming paper to
produce a pruned stellar catalogue (Groenewegen et al., 2001).  The
same techniques will also be applied to analyse the faint source
catalogue of the CDF-S which should also yield interesting results and
more insight regarding the source catalogues.

Finally, one should keep in mind that, unless follow-up spectroscopic
studies of the current samples are carried out, the numbers and
target list will remain tentative.

\begin{acknowledgements}

We would like to thank R. Ibata and B. Oppenheimer for providing us
with observed, very cool white dwarf spectra. We thank G. Chabrier and
I. Baraffe for providing their most recent model spectra of brown
dwarves, as well as D. K\"oster for giving us his most recent white
dwarf models. We also thank P. Bergeron for computing the expected
colour tracks of white dwarves in our filter system. Finally, we thank
D. Swayne from the AT\&T Labs, for providing and helping us install
the multi-dimension visualisation tool, $Xgobi$.

\end{acknowledgements}

\end{document}